\documentclass[fleqn,usenatbib]{mnras}
\pdfoutput=1

\usepackage[T1]{fontenc}
\usepackage{ae,aecompl}

\usepackage{siunitx}
\usepackage{graphicx}	
\usepackage{booktabs}
\usepackage{amsmath}	
\usepackage{amssymb}	
\usepackage{comment}

\def\Htwo{H$_{\rm 2}$}
\def\kms{km\,s$^{-1}$}
\def\LpUNIT{K km s$^{-1}$ pc$^{2}$}
\def\z{$z$\,}
\def\percc{cm$^{-3}$\,}
\def\plus{$^{+}$}

\def\mic{\,$\mu$m\,}
\def\lsun{\,L$_{\odot}$\,}
\def\Nplusa{[N{\sc ii}]\,205\mic}
\def\Nplusb{[N{\sc ii}]\,122\mic}
\def\Hplus{H{\sc ii} \,}
\def\Nplus{[N{\sc ii}] \,}
\def\Nplusplus{[N{\sc iii}] \,}

\title[Singly ionised nitrogen and multi-J CO study of the ``Red Radio Ring'' at  $z \sim 2.55$]{The ``Red Radio Ring'': Ionised and Molecular Gas in a Starburst/Active Galactic Nucleus at $z \sim 2.55$  }

\author[K. C.~Harrington et al.]{Kevin ~C. ~Harrington$^{1,2}$\thanks{E-mail: kharring@astro.uni-bonn.de}
A. Vishwas$^{3}$,
A. ~Wei{\ss}$^{4}$,
B.~Magnelli$^{1}$,
\newauthor
L.~Grassitelli$^{1}$,
M. Zaja\v{c}ek$^{4,5,6}$,
E. F. ~Jim{\'e}nez-Andrade$^{1,2}$,
\newauthor
T. K. D.~Leung$^{3,7}$,
F.~Bertoldi$^{1}$,
E. Romano-D{\'i}az$^{1}$
D.T.~Frayer$^{8}$,
\newauthor
P.~Kamieneski$^{10}$,
D.~Riechers$^{3,9}$,
G. J. Stacey$^{3}$,
M.S.~Yun$^{10}$,
Q.D.~Wang$^{10}$,\\
$^1$Argelander Institut f\"{u}r Astronomie, Auf dem H\"{u}gel 71, 53121 Bonn, 
Germany\\
$^2$International Max Planck Research School of Astronomy and Astrophysics at the Universities of Bonn and Cologne;\\
$^3$Department of Astronomy, Cornell University, Space Sciences Building, Ithaca, NY 14853, USA\\
$^4$Max-Planck-Institut f\"ur Radioastronomie (MPIfR), Auf dem H\"ugel 69, D-53121 Bonn, Germany\\
$^5$Center for Theoretical Physics, Polish Academy of Sciences, Al.  
Lotnikow 32/46, 02-668
Warsaw, Poland\\
$^6$I. Physikalisches Institut der Universit\"at zu K\"oln, Z\"ulpicher Strasse 77, D-50937 K\"oln, Germany \\
$^7$Center for Computational Astrophysics, Flatiron Institute, 162 Fifth 
 Avenue, New York, NY 10010, USA\\
$^8$Green Bank Observatory, 155 Observatory Rd., Green Bank, West Virginia 24944, USA\\
$^9$Max-Planck-Institut 
f\"{u}r Astronomie, K\"{o}nigstuhl 17, D-69117 Heidelberg, Germany\\
$^{10}$Department of Astronomy, University of Massachusetts, 619E Lederle Grad 
Research Tower, 710 N. Pleasant Street, Amherst, MA 01003, USA\\
}

\date{Accepted XXX. Received YYY; in original form ZZZ}

\pubyear{2018}
\hypersetup{draft}
\begin{document}
\label{firstpage}
\pagerange{\pageref{firstpage}--\pageref{lastpage}}
\maketitle

\begin{abstract}

We report the detection of the far-infrared (FIR) fine-structure line of singly ionised nitrogen, \Nplusa,  within the peak epoch of galaxy assembly, from a strongly lensed galaxy, hereafter ``The Red Radio Ring'';  the RRR, at z = 2.55. We combine new observations of the ground-state and mid-J transitions of CO (J$_{\rm up} =$ 1,5,8), and the FIR spectral energy distribution (SED), to explore the multi-phase interstellar medium (ISM) properties of the RRR. All line profiles suggest that the HII regions, traced by \Nplusa, and the (diffuse and dense) molecular gas, traced by the CO, are co-spatial when averaged over kpc-sized regions. Using its mid-IR-to-millimetre (mm) SED, we derive a non-negligible dust attenuation of the \Nplusa line emission. Assuming a uniform dust screen approximation results a mean molecular gas column density $> 10^{24}$\, cm$^{-2}$, with a molecular gas-to-dust mass ratio of 100. It is clear that dust attenuation corrections should be accounted for when studying FIR fine-structure lines in such systems. The attenuation corrected ratio of $L_{\rm NII205} / L_{\rm IR(8-1000\mu m)} = 2.7 \times 10^{-4}$ is consistent with the dispersion of local and $z >$ 4 SFGs. We find that the lower-limit, \Nplusa -based star-formation rate (SFR) is less than the IR-derived SFR by a factor of four. Finally, the dust SED, CO line SED and $L_{\rm NII205}$ line-to-IR luminosity ratio of the RRR is consistent with a starburst-powered ISM.
\end{abstract}




\section{Introduction}

Observational evidence reveals a synchronous peak, around $z \sim 2$, in both the cosmic co-moving star-formation rate (SFR) and super massive black hole accretion rate density  \citep[see e.g. ][]{madau14,hickox18}. Understanding this apparent co-evolution between active galactic nuclei (AGN) and star formation (SF) demands a deeper characterisation of the interstellar medium (ISM) in galaxies, such as the dynamics and spatial distribution of gas arising from different phases, as well as the relationship of ionised, molecular and stellar surface mass densities and their role in SF processes. Substantial theoretical work \citep{ dallavecchia08, scannapieco12, rosdahl17} has also progressed in simulating the complex effects of black-hole, thermal, and kinetic feedback processes, while observations of  ISM properties derived from a broad-band coverage are still required to form a complete impression of a galaxy that has both AGN and SF activity \citep{cicone14, cicone15, cicone18}. High-$z$ star-forming galaxies (SFGs) at \z $\sim 1-3$ typically show an increase in the molecular gas-to-stellar mass fractions (up to 50\% or greater) \citep[e.g.][]{tacconi10,Tacconi2018}. The spatial extent of SF within high-\z SFGs can often exist out to large radii \citep[$\sim$ 2-10 kpc e.g.][]{hd10,brisbin15,magdis16,elbaz18}, exceeding the 0.1-1 kpc nuclear starburst (SB) regions of local (Ultra)Luminous InfraRed Galaxies \citep[LIRGs have $10^{10} < L_{\rm IR(8-1000\mu m)} < 10^{11}$ \lsun and ULIRGs have $L_{\rm IR(8-1000\mu m)} > 10^{12}$ \lsun][]{sanders96, solomon97,sv05}. Therefore global properties derived from measurements of the ionised and molecular ISM are needed to account for the total emission corresponding to the kpc-scale areas encompassed by high-\z systems.

Studying the gas-rich, dusty star-forming galaxies (DSFGs) at $z > 1$ has largely focused on measurements of the molecular gas content via one or two CO lines (typically J$_{\rm up} \le 5$), and also the long-wavelength dust continuum, to understand the star-forming ISM, the total molecular gas mass and overall efficiency of SF \citep[e.g.][]{genzel10,scoville14,Schinnerer2016,scoville16,scoville17,harrington18, leung19}. The ionised ISM, however, has been largely unexplored at high-$z$, and therefore the complete picture of multi-phase gas processes required to disentangle the nature of SF in galaxies are poorly constrained. Far-IR fine-structure lines (FSLs) offer an additional probe of \Hplus regions in obscured sites of SF, as they are less susceptible to dust attenuation when compared to optical or mid-IR lines \citep[see ][]{fernont16, diazsantos17}. This motivates the use of these far-IR FSLs as powerful line diagnostics of the evolving ISM at high-$z$ \citep{maoilino05,Maiolino2009,  ferkinhoff10,ferkinhoff11,riechers14,zavala18,zhang18,lamarche18,marrone18,vishwas18,zanella18}. Unfortunately, the atmospheric coverage of many important mid-/far-IR FSLs makes observations difficult to execute, if not impossible to observe from the ground. 

The nitrogen atom has an ionisation energy E$_{i,\rm N}$= 14.53\,eV, and is therefore typically present with singly ionised hydrogen; E$_{i,\rm  H}$= 13.6\,eV. The fine-structure splitting of the ground-state leads to two transitions at 121.898\,$\mu$m and 205.178$\,\mu$m; \Nplusb and \Nplusa, respectively\footnote{The ground state ($^{3}P_{0}$) fine-structure splitting arises due to the unpaired electrons in the nitrogen atom. The $^{3}P_{2}$ and $^{3}P_{1}$ levels are only about 188\,K and 70\,K above ground, respectively. }. In order to characterise the global ionised ISM properties, the low ionisation energy requirement of the far-IR \Nplus emission lines makes them unique tracers of the low-excitation, warm ionised gas associated with \Hplus regions and the ambient interstellar radiation field of the ISM. The physical and chemical evolution of the global ISM is influenced by supernova explosions and high mass-loss rates dispelled by stellar winds from massive OB and Wolf-Rayet type stars \citep[e.g.][]{mckee97,crowther07,puls08}. These, together with efficient rotational mixing within massive
stars \citep{maeder00,brott11a,ekstrom12}, can quickly expose the products of stellar nucleosynthesis at 
the surface, thereby injecting
substantial quantities of nitrogen into the ISM within a timescale of $\sim$10s Myr \citep{maeder00,stanway18}.

The \Nplus emission lines were first observed in the Milky Way by the COBE FIRAS spectrometre \citep{bennett94}, followed closely by KAO observations of the Galactic \Hplus region G333.6-0.2 \citep{colgan93}. The \Nplusa line is also observable at rest velocities from the ground-based observatories at exceptional sites. Using the SPIFI spectrometer on the AST/RO telescope at South Pole, \citet[][]{oberst06,oberst11} mapped the \Nplusa line from the Carina Nebula and compared it with ISO LWS \Nplusb line maps to show the [NII] line originated from a low density ($n_{e^{-}} \sim 28$ \percc) ionized medium. High spatial-resolution, large-scale imaging of the Galactic plane were enabled by the sensitive PACS and SPIRE spectrometer on-board the \textit{Herschel Space Observatory} \citep{goldsmith15}, and demonstrated that most of the \Nplus line arises from extended, low density ($n_{\rm e^{-}}$ $\sim$ 10 to 50 cm$^{-3}$) \Hplus regions. Other efforts to use \Nplus to derive average electron densities  have been made in a range of local galaxies ($n_{\rm e^{-}}$ $\sim$  20-100 cm$^{-3}$), for instance: M51 and Centaurs A, \citep{parkin13}, Ultra-luminous Infrared Galaxies ULIRGs \citep[HERUS sample; ][]{farrah13}, Dwarf galaxies \citep{cormier15}, KINGFISH galaxies \citep{hc16} and other SFGs \citep{lu17}. 

At high-$z$, observations of the \Nplusa line is largely limited to $z > 3.9$, where the line is red-shifted to wavelengths longer than 1mm, making ground based observations possible due to the more transmissive and stable atmosphere, with lower receiver noise temperatures. The current sample where this emission line is detected consists of twelve highly star-forming galaxies \citep{decarli12,decarli14,combes12,rawle14,nagao12,bethermin16,pavesi16,pavesi18b,pavesi18c,lu18}, and there are at least five additional non-detections \citep[see][]{walter09,riechers13}.

In this paper we report new spatially unresolved line detections from ``The Red Radio Ring'' (hereafter: the \textit{RRR}) of \Nplusa line emission with the APEX telescope, complemented by CO(1-0), CO(5-4), and CO(8-7) line detections from the \textit{Green Bank Telescope} (GBT) and IRAM 30m telescope. The \Nplusa line detection at the redshift, $z \sim 2.55$, in the \textit{RRR} begins to bridge the gap between local detections and those at $z > 4$. \\

We structure the paper as follows, in \S 2 we provide a brief outline of the nature of the galaxy presented in this study. We describe the \Nplus and CO observations in \S 3, and then present the results in \S 4. In \S 5 we discuss the \Nplus derived SFR and the possibility for a co-eval AGN/SB, followed by our conclusions and outlook in \S 6. Throughout this paper we take for a point of reference a flat $\Lambda$ CDM cosmology with ${\rm H_{0} = 69.6 \, km s^{-1} Mpc^{-1} 
}$ with ${\rm \Omega _{m} = 0.286}$, and ${\rm \Omega _{\Lambda} = 1- \Omega _{m}}$ \citep{bennett14}. Throughout the text, we use a magnification factor, $\mu =$ 15, to report the intrinsic source properties unless otherwise noted. This value is derived from lens models using the highest spatial resolution data available for this source, i.e. $\mu = 14.7 \pm 0.3$, \citep{geach18}, and is consistent with 
other work \citep[][; Kamieneski et al. 2019, in prep.]{rivera18}. The relative magnification factor, however, can change depending on the source plane distribution of every line and continuum tracer at varying rest frequencies.

\section{The Red Radio Ring}
the \textit{RRR} was discovered by four independent teams: \textbf{(i)} the citizen science program \textit{SpaceWarps} \citep{marshall16} in a search for gravitational lensing features within deep (iJKs band) CFHT images in the \textit{Herschel}-Stripe82 field \citep{geach15}; \textbf{(ii)} \citet{harrington16} identified this source after cross-matching \textit{Herschel}-SPIRE and \textit{Planck} images at 350\micron \, in order to identify strongly lensed, DSFGs \citep{negrello10,fu12,planck26,wardlow13,canameras15} and further confirmed with follow-up CO and millimetre dust continuum observations with the \textit{Large Millimeter Telescope}; \textbf{(iii)} \citet{nayyeri16} present a similar selection of candidate lensed, DSFGs as \citet{harrington16}, but with SPIRE 500\micron \, images instead; and \textbf{(iv)} \citet{su17} identified the \textit{RRR} as the brightest DSFG candidate (referred to as ACTJ0210+0016) in the 148, 218, and 278 GHz maps from the Atacama Cosmology Telescope (\textit{ACT}), and presented follow-up CO(1-0) line observations with \textit{Green Bank Telescope}/Zpectrometer.

the \textit{RRR} is a strongly lensed radio-AGN/DSFG hybrid galaxy, magnified by a massive, foreground elliptical galaxy and a satellite companion at $z = 0.2019$ \citep{geach15}. The 1.4 GHz eMERLIN imaging ($\theta \sim 0.35"$) revealed compact radio emission $<$250pc in the lens reconstructed source-plane image. The intrinsic specific radio luminosity $L_{\rm 1.4 GHz} \approx 10^{25}\, {\rm W\,Hz^{-1}}$ suggests a radio-mode AGN \citep{geach15}. The wavelength corresponding to the peak line flux of the asymmetric low-J CO line profile corresponds to a redshift, z$\sim$2.553 for the \textit{RRR} \citep{harrington16,su17}. Detailed strong lens modeling of the CO(3-2) emission resolved by NOEMA ($\theta \sim$ 0.75$\times$1.5'') suggests that the observed molecular gas emission arises from a rotating disk spread over $\sim$3\,kpc in the source-plane \citep{rivera18}. The source-plane reconstruction of the CO(4-3) line emission ($\theta \sim$ 0.25'') by \citet{geach18} provides evidence that the growth of the AGN is co-eval with the rapid SF. The molecular gas may dominate the galactic potential within these three kpc, which is further supported by \citet{rivera18}.


\begin{figure*}

	\includegraphics[width=0.99\textwidth]{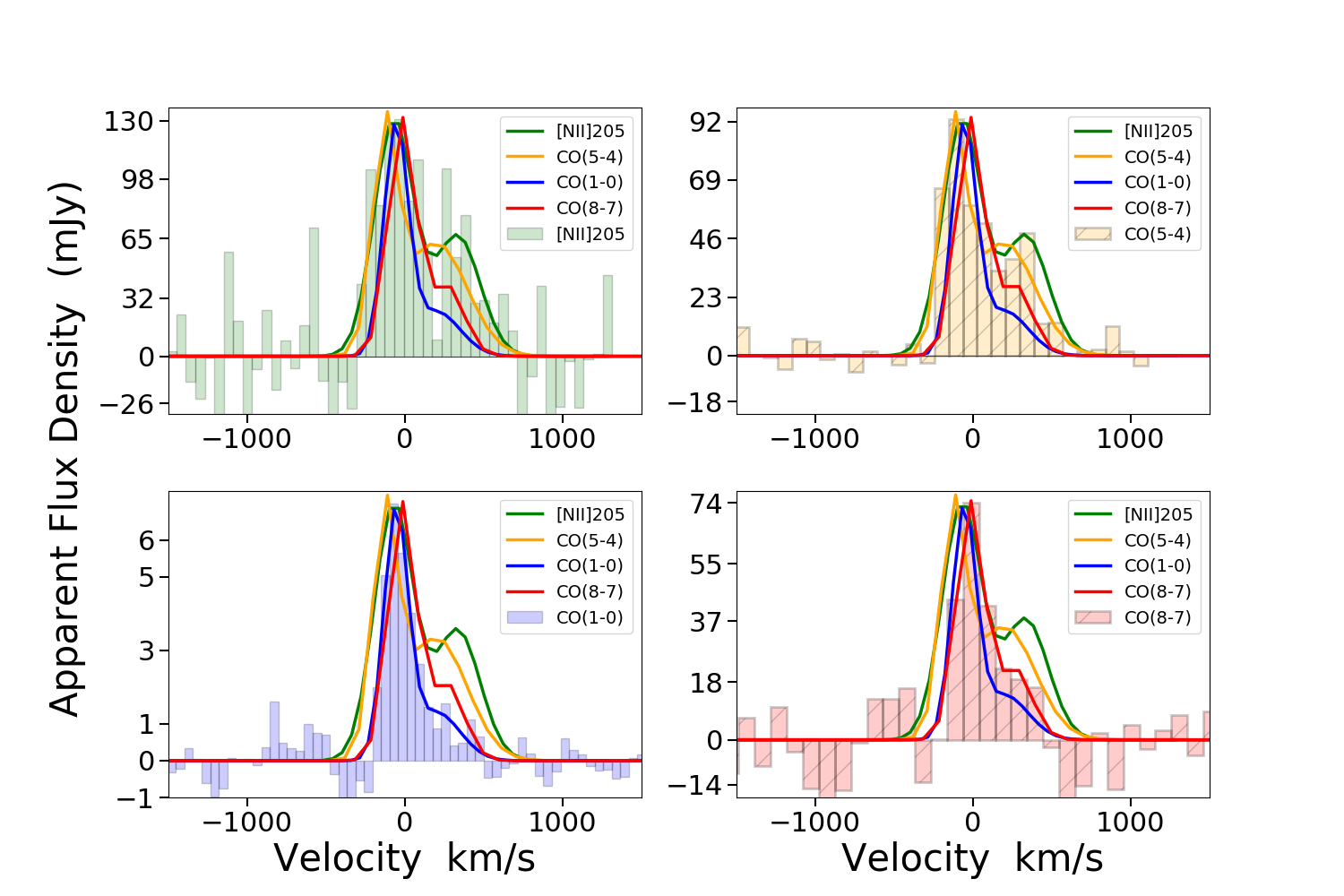}
    \caption{The spectra and two-component Gaussian fits for the \Nplusa (green; top left), CO(1-0) (blue; bottom left), CO(5-4) (orange; top right) and CO(8-7) (red; bottom right) lines. To aid comparisons among all line profiles, the best-fit Gaussian models have been re-scaled to the observed peak of each spectrum within each panel. The zero-point velocity is determined using $z = 2.553$.}
    \label{fig:overlays}
\end{figure*}

\section{Observations}
\subsection{GBT}
The CO(1-0) line emission was observed using the Ka-band receiver on the GBT. Observations (GBT/17B-305;
PI: K. Harrington) took place on October 22, 2017, under stable atmospheric conditions. We used the standard SubBeamNod procedure between the 8-m subreflector and the main dish, with 4 min integrations per scan. Pointing and focus were performed frequently before
the SubBeamNod integrations. The backend spectrometer, VEGAS, was used to record the data from the Ka-band receiver, tuned to the expected CO(1-0) line frequency (in low-resolution, 1.5 GHz
bandwidth mode; $\theta \sim 23"$). Subsequent data reduction was performed using GBTIDL \citep{marganian13}. All On-Off measurements were corrected for the
atmospheric attenuation and afterwards treated in the same manner as in  \citet{harrington18}. We smoothed all spectra to 50 \kms channel resolution after averaging all low-order baseline subtracted spectra. The resulting on-source integration time was 1.25h. Flux accuracy
was checked with the standard source Uranus and pointing stability with J0841+7053, J1310+3220, J1331+3030 and
J1642+3948. We adopt a 25\% uncertainty on the integrated line properties for systematic effects with the GBT \citep[see][]{harrington18,frayer18}. 

\subsection{APEX}
To observe the \Nplusa emission line we used the FLASH\plus 460L single polarization receiver on the Atacama Pathfinder EXperiment (APEX) 12m telescope \citep{guesten06}. We used Max Planck Society observing time between 24 May and 17 July, 2018 (Pr. M-0101.F-9503A-2018; PI: Harrington), totaling 384 minutes of integration ($\theta \sim 15"$). FLASH \citep{heyminck06} is a 2 side-band (SB) dual-frequency heterodyne receiver with orthogonal linear polarizations, one for each of the 345 GHz and 460 GHz atmospheric windows. The FLASH observations were performed in good weather conditions, with precipitable water vapor < 1.5mm. Observations used standard wobbler switching with a chopping rate of 1.5 Hz, and an azimuthal throw offset of 30''. Each scan consisted of a hot/sky/cold calibration 600'' off-source, followed by 12 subscans of 20s per on-source integration time. Focus checks were performed regularly (every 3-5h), whereas pointing checks on a strong line/continuum source (e.g. Jupiter or nearby star) were performed roughly every 1-2h and yield a pointing accuracy within 2-3". To record the data we used the MPIfR eXtended bandwidth Fast Fourier Transform spectrometers \citep[FFTS;][]{klein06} with a 2$\times$ 2.5 GHz bandwidth for each of the upper and lower receiver sidebands of spectra the FLASH receiver. All scans were reduced and analysed using the CLASS and GREG packages within the GILDAS\footnote{Software information can be found at: http://www.iram.fr/IRAMFR/GILDAS } software distribution. Each scan was smoothed to $\sim90$ km s$^{-1}$ channel resolution and assessed by eye after a 1st order baseline polynomial subtraction (line-free channels). Only about 10\% of the scans were removed for each set of spectra before co-adding the rms-weighted spectrum. We adopt an absolute uncertainty of 25\% for all derived line properties to account for the variations in systematic behavior of the APEX observations at higher frequencies (e.g. atmospheric stability, baseline subtraction, pointing/focus corrections).

\subsection{IRAM 30m}
Observations with the IRAM 30m telescope took place across two observing semesters: Pr. 187-16 and Pr. 170-17 (PI: K. Harrington), starting on January 29th, 2017 we observed the CO(5-4) emission line in average weather conditions ($\tau_{\rm \nu obs} = 0.5-0.8$) for 30 minutes of integration. Subsequent observations were in excellent observing conditions ($\tau_{\rm \nu obs} < 0.04-0.2$) on December 13, 2017 for roughly 35 minutes integration to detect the CO(8-7) emission line.
We used the E150 and E230 observing bands of the EMIR receiver, and utilised two backends: both the WIde-band Line Multiple Auto-correlator (WILMA) and the fast Fourier Transform Spectrometre (FTS200). Our observing mode consisted of a single EMIR band, capturing the dual polarization, 16 GHz bandwidth of the lower inner and lower outer (LI+LO), and upper inner and upper outer (UI+UO) sidebands with respect to the LO tuning frequency. To overcome the variable atmospheric conditions, we used the wobbler switching observing mode to perform offset throws of 40\arcsec every second.  Each wobbler switching mode procedure includes three, 5 minute integrations (i.e. twelve 25-s subscans). Frequent focus and pointing checks were assessed (e.g. Uranus, Venus, J1226+023, J1418+546) every 1.5 to 2hr, with azimuth and elevation pointing offsets typically within 3\arcsec. The IRAM 30m beam sizes at the observed CO(5-4) and CO(8-7) line frequency are $\theta \sim 15"$ and  $\theta \sim 10"$, respectively. The absolute uncertainty we adopt for the derived line properties from the IRAM 30m observations is 20\% based on the dispersion of flux densities observed in pointing sources from ongoing monitoring at the telescope. All scans were reduced using GILDAS, smoothed to $\sim 50$ \, km s$^{-1}$ channel resolution before being co-added.

\section{Results}

\subsection{Intrinsic Line Properties}
The observed \Nplusa emission line peaks at $\nu^{\rm peak}_{\rm obs}=411.2485$ GHz. We integrate the full line profile to derive a total velocity integrated flux density of 4.3 $\pm$ 1.1 Jy \kms (using an antenna gain factor of 52.3 Jy/K). The CO(1-0) ($\nu^{\rm peak}_{\rm obs}\,=32.4432\pm0.0001$ GHz) has a measured integrated flux of 0.18 $\pm$ 0.04 Jy \kms (antenna gain factor of 0.7 Jy/K). This is consistent with, albeit slightly higher than, the Zpectrometer measurement (0.11$\pm0.03$ Jy \kms) of \citet{su17}. The velocity integrated flux intensities for the CO(5-4) ($\nu^{\rm peak}_{\rm obs}=162.212$ GHz) and CO(8-7) ($\nu^{\rm peak}_{\rm obs}=259.484$ GHz) emission lines are 1.96 $\pm $0.3 and 1.37 $\pm $0.27 Jy \kms, using antenna gain factors of 6.69 and 8.38 Jy/K, respectively. We report in Tab.~\ref{table2} the line luminosity (in L$_{\odot}$) and spatially integrated source brightness temperature (in \LpUNIT) following \citet{carilli13}. We note that the peak line intensity frequencies are all consistent with $z = 2.5535 \pm 0.0006$.

Asymmetric line profiles are observed in all the high S/N (S$_{\rm peak}$/N$_{\rm rms} > 10$) line detections (CO 1-0,\,5-4,\,8-7; \Nplusa), therefore we fit two 1-D Gaussians to the line shapes to compare their respective full-width-at-half-maximum (FWHM), centroids and amplitudes. The best-fit models are overlaid on the CO and \Nplusa spectra in Fig.~\ref{fig:overlays}, while the best-fit parameters are listed in Tab.~\ref{table2}, together with the CO(3-2)\footnote{As noted in \citet{rivera18}, the CO(3-2) line flux for the LMT detection presented in \citet{harrington16} is unfortunately incorrect due to the early commissioning period and calibration uncertainties.} and CO(4-3) velocity integrated line flux densities from \citet{rivera18,geach18}. 

The line centroid and FWHM of the \Nplusa emission line are consistent with the observed CO(1-0) (tracing the total molecular gas mass), and the more highly excited, J$_{\rm up} > 3$, CO lines. In all lines, the red component is offset by about 250-450 \kms from the blue component. Differential lensing may yield differences in measured line ratios \citep{serj12}. However, without higher angular resolution observations for each transition, we assume the magnification factor does not change for each of the observed lines, such that the low-density diffuse \Hplus regions traced by \Nplusa and the molecular gas traced by CO are considered to be co-spatial when averaged across kpc scales.

\begin{table*}
\caption{Best-fit Gaussian Models and Line Properties}
\begin{tabular}{lllllll}
\hline
\multicolumn{1}{|c|}{} & \multicolumn{1}{l|}{{[}NII{]}205$\mu$m} & \multicolumn{1}{l|}{CO(1-0)} & \multicolumn{1}{l|}{CO(3-2)$^b$} & \multicolumn{1}{l|}{CO(4-3)$^c$} & \multicolumn{1}{l|}{CO(5-4)} & \multicolumn{1}{l|}{CO(8-7)} \\ \hline
Redshift, $z$ (peak) & 2.55308 (0.0004) & 2.55300 (0.0004) & 2.5529 (0.00011) & 2.5543 (0.0002) & 2.5525 (0.0002) & 2.55243 (0.0003) \\ \hline
\multicolumn{7}{|c|}{Total Intrinsic Line Properties$^a$:} \\ \hline
$S_{\nu} \Delta V$ {[}Jy km s$^{-1}${]} & 4.3 (1.1) & 0.18 (0.04) & 1.38 (0.28) & 1.62 (0.32) & 1.96 (0.3) & 1.37 (0.27) \\
$L' {[}10^{10}$ K km s$^{-1}$ pc$^{2}${]} & 0.84 (0.21) & 3.67 (0.92) & 4.8 (1.2) & 3.2 (0.8) & 2.5 (0.6) & 0.67 (0.19) \\
$L_{\rm line}${[}10$^{8}$ L$_{\odot}${]} & 8.4 (2.0) & 0.02 (4.3e-03) & 0.64 (0.02) & 1.0 (0.25) & 1.5 (0.38) & 1.6 (0.42) \\ \hline
\multicolumn{7}{|c|}{Component A (peak):} \\ \hline
FWHM (km s$^{-1}$) & 293 (76) & 179 (23) & - & - & 165 (22) & 154 (50) \\
Amplitude (mJy) & 8.7 (1.3) & 0.44 (0.07) & - & - & 5.5 (0.7) & 5.1 (2.0) \\
Center (km s$^{-1}$) & -69 (34) & -55 (10) & - & - & -37 (6) & -20 (17) \\ \hline
\multicolumn{7}{|c|}{Component B:} \\ \hline
FWHM (km s$^{-1}$) & 337 (170) & 352 (230) & - & - & 480 (116) & 337 (353) \\
Amplitude (mJy) & 4.5 (1.3) & 0.09 (0.023) & - & - & 2.10 (0.25) & 1.5 (0.6) \\
Center (km s$^{-1}$) & 334 (72) & 200 (118) & - & - & 240 (60) & 200 (188) \\
 &  &  &  &  &  & \\
 \hline
\end{tabular} 
\\
The velocities are measured with respect to $z = 2.553$, i.e. the peak velocity component. $^{a}$ Measured line properties corrected for magnification ($\mu =15$). The systematic errors are listed in parenthesis for the velocity-integrated flux density and derived total line luminosities. The parenthessis associated with the FWHM, centroid, and amplitude for Gaussian components A and B are based on the residual errors to the model fit. $^{b}$ \citet{su17,rivera18} and $^{c}$\citet{geach18}, corrected for the cosmology used throughout this paper. \citet{geach18} report the redshift based on the mid-point full-width-at-zero-intensity of the observed CO(4-3) transition.
\label{table2}
\end{table*}
\subsection{Far-IR Spectral Energy Distribution}

\begin{figure*}

	\includegraphics[width=0.95\textwidth]{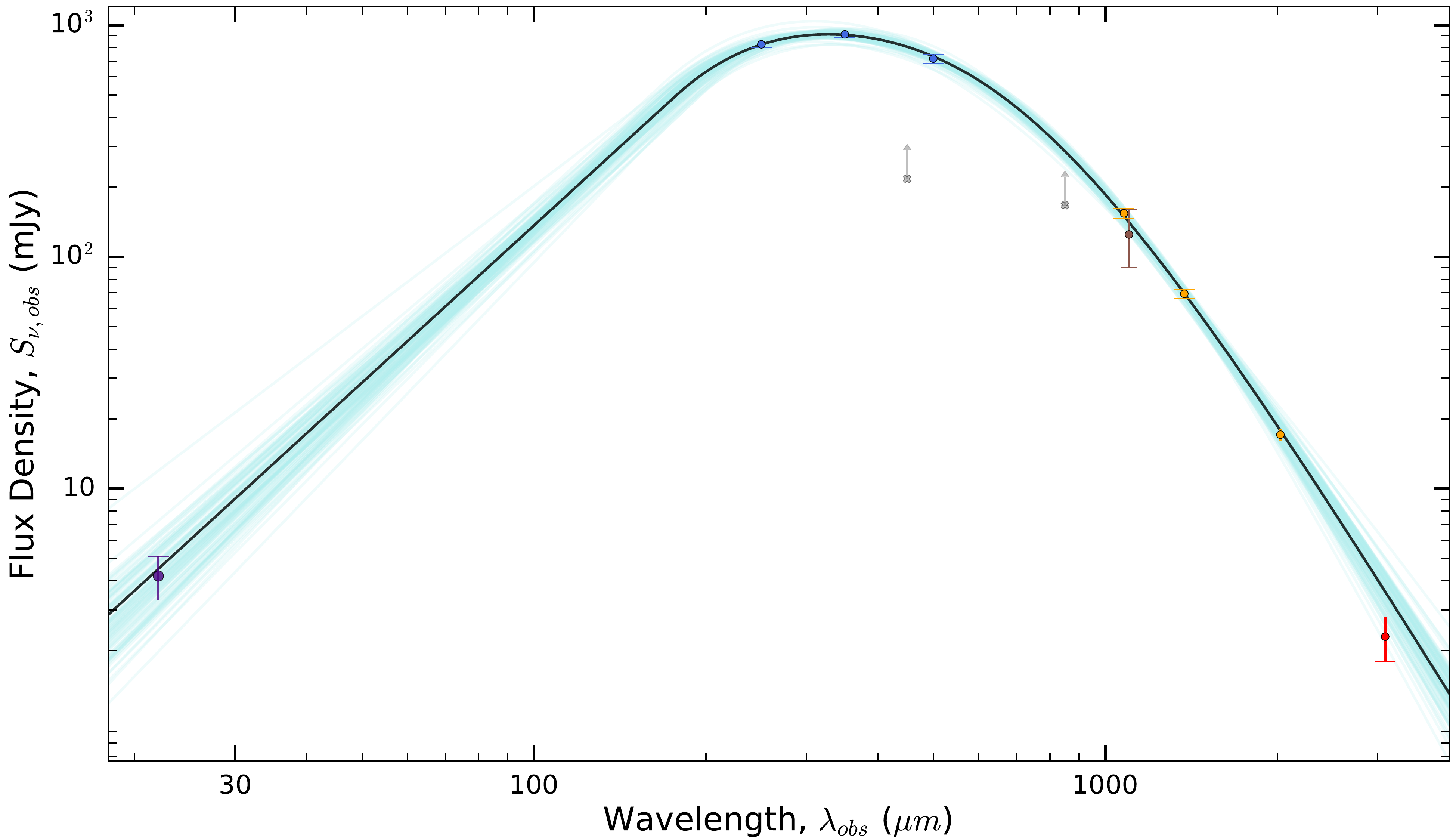}
    \caption{The best-fit modified blackbody SED model (black line) for the \textit{RRR}. We also show multiple iterations of the models created by sampling the parameter space for the modified blackbody (cyan) that are representative of the degeneracies in the parameter space. Data included for the SED fit exercise is shown as colored circles with corresponding error bars: (indigo) WISE/W4, (blue) Herschel/SPIRE, (yellow) ACT \citep{su17} and (red) CARMA \citep{su17}. For completeness, we show data that is not included for the SED fit - (gray cross as lower limits) measurements from SCUBA-2 presented in \citet{geach15} and  (brown circle) is the average of the two AzTEC/LMT measurements from \citet{geach15} and \citet{harrington16}.}
    \label{fig:sed}
\end{figure*}

\begin{figure*}

	\includegraphics[width=0.95\textwidth]{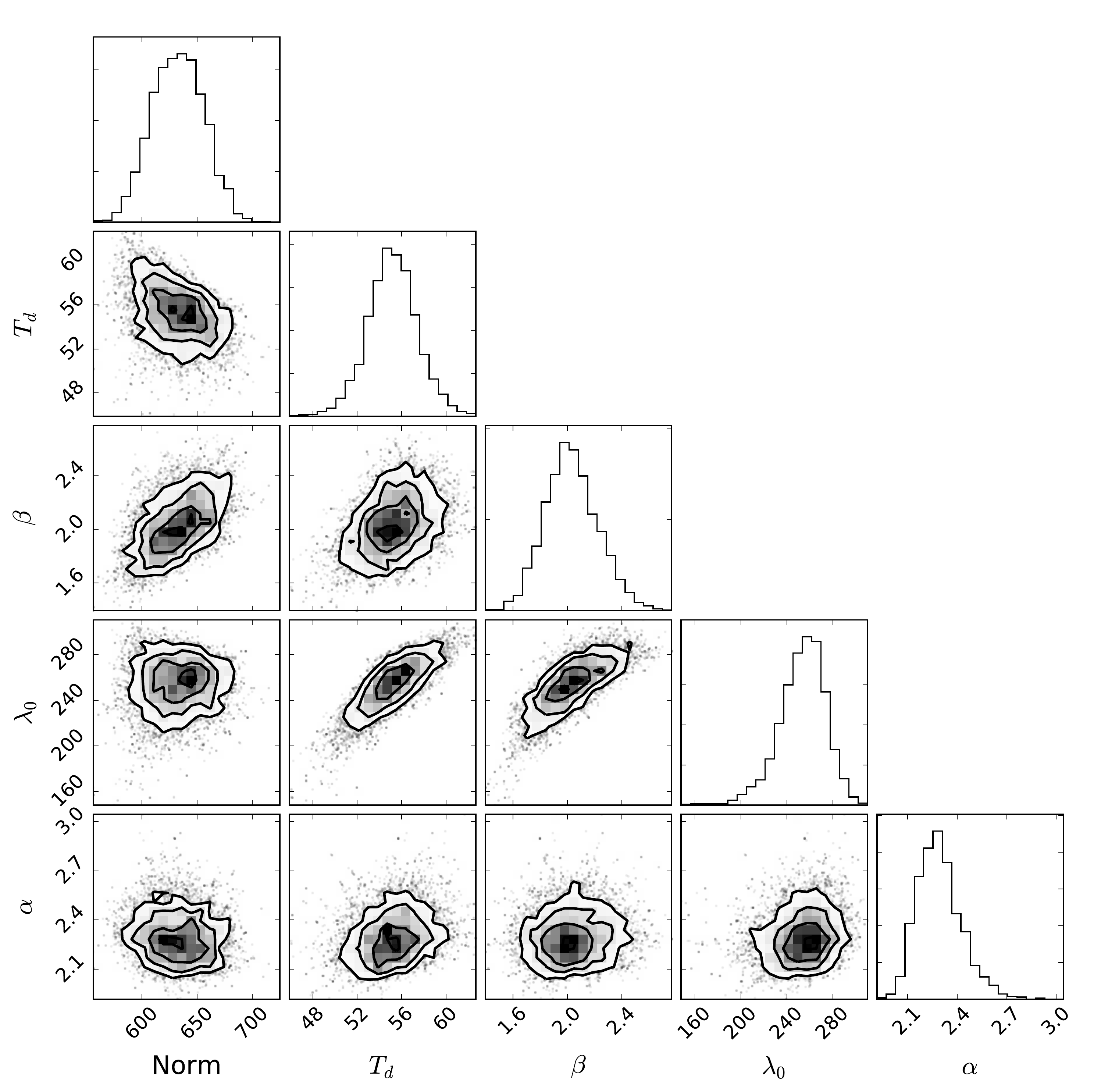}
    \caption{Posterior probability distribution for the SED model parameters: the value at which the dust opacity reaches unity is at rest-frame wavelength of $\lambda_{0}$, dust emissivity index, $\beta$, a single component dust temperature, $T_{\rm d}$, and the Wien-side power law slope for SFGs, $\alpha$. }
    \label{fig:sed_posterior}
\end{figure*}

\begin{table}
\caption{Observed mid-IR to mm photometry for the \textit{RRR}}
\label{tab:phot_tab}
\resizebox{0.5\textwidth}{!}{%
\begin{tabular}{@{}lllll@{}}
\toprule
Wavelength (\micron) & Flux Density (mJy) & Instrument        &  &  \\ \midrule
22         & 4.2$\pm$0.9  & WISE/W4           &  &  \\
250        & 880$\pm$27   & Herschel/SPIRE    &  &  \\
350        & 991$\pm$30   & Herschel/SPIRE    &  &  \\
500        & 773$\pm$33   & Herschel/SPIRE    &  &  \\
850        & 167$\pm$4    & JCMT/SCUBA (G15)  &  &  \\
1078.4     & 154$\pm$8    & ACT (278GHz, S17) &  &  \\
1100       & 95.5$\pm$6   & LMT/AzTEC (G15)   &  &  \\
1100       & 145$\pm$15   & LMT/AzTEC (H16)   &  &  \\
1375.2     & 69$\pm$3     & ACT (218GHz, S17) &  &  \\
2025.6     & 17$\pm$2     & ACT (148GHz, S17) &  &  \\
3090.6     & 2.3$\pm$0.5  & CARMA (S17)       &  & \\
\hline
\end{tabular}%
}\\
\citet[][G15]{geach15}, \citet[][H16]{harrington16}, \citet[][S17]{su17}.
\end{table}

Using data from the literature and various telescope archives, we compiled multi-band photometry tracing emission from the \textit{RRR} in the (observed-frame) mid-infrared to mm-wavelengths \citep{geach15, harrington16, Schulz17, su17}. We fit the observed SED with a single temperature modified blackbody (MBB) model combined with a Wien-side power-law slope, denoted as $\alpha$, of which a value of $\alpha \sim $2 is characteristic for SFGs \citep[e.g.,][]{casey12}. If an AGN torus is contributing a hot dust component, the MIR would show an excess compared to the power law slope for a normal SFG \citep[e.g. the WISE 'Hot DOGS'][]{tsai15}.\\

We retrieved the SPIRE photometer measurements from the SPIRE point source catalog \citep{Schulz17}. We report the estimated uncertainties due to confusion rather than the systematic and statistical errors (which are $<$2\%) as the SPIRE beam is large (18-35'') and the diameter of the radio Einstein ring is roughly 5'' \citep{geach15}. We find that the flux density measured with LMT/AzTEC varies by 50\% between the two observations by \citet{geach15} and \citet[][see footnote 3 above]{harrington16}. For the MBB fit, we use the average value of these measurements, with an uncertainty that encompasses the range of values reported, i.e., 125$\pm$35\,mJy. We note that \citet{geach15} reported that the 450\micron \,flux density measured with SCUBA was a factor of three smaller than that measured by SPIRE and hence this SCUBA measurement not included here. In comparison with the ACT 278\,GHz measurement, we find that the 850\mic SCUBA flux density, though a high significance detection, is most likely an underestimate, perhaps due to absolute flux calibration. We therefore ignore the SCUBA 850\mic data point while performing the model fitting.  
 
Fig.~\ref{fig:sed} shows the best-fit SED model of the \textit{RRR}. In the following we quote the best-fit and the uncertainties based on the 16$^{th}$, 50$^{th}$ and 84$^{th}$ percentiles of the samples in the marginalized distributions for each of the parameters (see Fig.~\ref{fig:sed_posterior}). We find the dust opacity reaches unity at rest-frame wavelength of $\lambda_{0}$ = 254$^{\rm +17}_{\rm -18}$ \micron, with a dust emissivity index $\beta$=2.0$^{\rm +0.17}_{\rm -0.17}$ and a dust temperature, $T_{\rm d} = 55^{\rm +2.3}_{\rm -2.2}$ K, Wien-side power-law slope $\alpha = 2.27_{\rm -0.11}^{\rm +0.14}$. The total \textit{apparent} infrared luminosity  (IR; 8 - 1000\micron), $\mu L_{\rm IR} = $ 21.9$^{\rm +1.0}_{\rm -0.89}\times$10$^{13}$\lsun, before correcting for the magnification factor, $\mu$. The \textit{apparent} far-infrared luminosity (FIR; 40-120\micron), is $\mu L_{\rm FIR} = $ 12.3$^{\rm +0.41}_{\rm -0.43}\times$10$^{13}$\lsun. The value of $L_{\rm IR}$/$L_{\rm FIR}$ is consistent with normal star-forming systems, i.e. $L_{\rm IR}$/$L_{\rm FIR} \approx 1.5-2$ \citep[][; see e.g. Leung et al. 2019b, submitted]{dale01}. Thus, the observed dust SED does not show strong signs of an AGN influence, e.g. no bright WISE/W4 counterpart. This suggests i.) that the compact radio-AGN, revealed by bright radio emission with a steep radio synchrotron slope of $\alpha_{\rm radio} = -1.1$ \citep{geach15}, does not significantly affect the overall IR luminosity of the \textit{RRR}, or ii.) there is extreme dust obscuration of an AGN. Its intrinsic SFR can thus be estimated using $\mu = 15$ \citep{geach18} and the standard calibration of the total IR to SFR, with \citep[ SFR$_{\rm IR} = 1.7\times 10^{-10} L_{\rm IR}$ \,  M$_{\rm \odot }$ yr$^{-1}$; ][]{kennicutt98}. We find the SFR$_{\rm IR} =$ 2482 $\pm$ 992 M$_{\rm \odot}$ yr$^{-1}$, taking into account the total error propagation for the average best-fit relative uncertainty on the IR luminosity ($\sim $3\%), and the systematic errors for the photometric data points used in the modeling ($\sim $37\%; Tab.~\ref{tab:phot_tab}). 
 
\begin{table*}
\caption{The intrinsic (lensing-corrected) properties of the \textit{RRR}}
\begin{tabular}{cccccc}
\hline
\multicolumn{1}{|c|}{$L_{\rm IR (8-1000\mu m)}$} & \multicolumn{1}{c|}{$L_{\rm FIR (40-120\mu m)}$} & \multicolumn{1}{c|}{SFR$_{\rm IR}$} & \multicolumn{1}{c|}{SFR$_{\rm [NII]205}$} & \multicolumn{1}{c|}{$M_{\rm H_{2}}$} & \multicolumn{1}{c|}{$M_{\rm min}$(H$^{+}$)} \\ \hline
\multicolumn{1}{|c|}{{[}10$^{13} L_{\odot}${]}} & \multicolumn{1}{c|}{{[}10$^{13} L_{\odot}${]}} & \multicolumn{1}{c|}{M$_{\odot}$ yr$^{-1}$} & \multicolumn{1}{c|}{\begin{tabular}[c]{@{}c@{}}M$_{\odot}$ yr$^{-1}$\end{tabular}} & \multicolumn{1}{c|}{10$^{10}$ M$_{\odot}$} & \multicolumn{1}{c|}{10$^{10}$ M$_{\odot}$} \\ \hline
\hline
1.46$_{-0.06}^{0.07}$ & 0.82$_{-0.03}^{0.04}$ & 2482 (992) & 621$^{b}$ & 3.67 (0.9) & 0.89\\
\hline
\end{tabular}
\\
The reported values can be converted back to the apparent values by multiplying the average lensing magnification factor of $\mu$=15 \citep{geach18}. The intrinsic SFR and $M({\rm H^{+}}$) are derived from the \Nplusa emission line, and are corrected by factor of 4.67 to account for the derived attenuation assuming a uniform dust screen approximation. The total and far-IR derived luminosities are derived from the dust SED modeling of the photometry from \citet{harrington16,su17,rivera18,geach15,geach18}.
\label{table1}
\end{table*}
\subsubsection{Effects of Dust Attenuation}
\begin{figure}

	\includegraphics[width=0.48\textwidth]{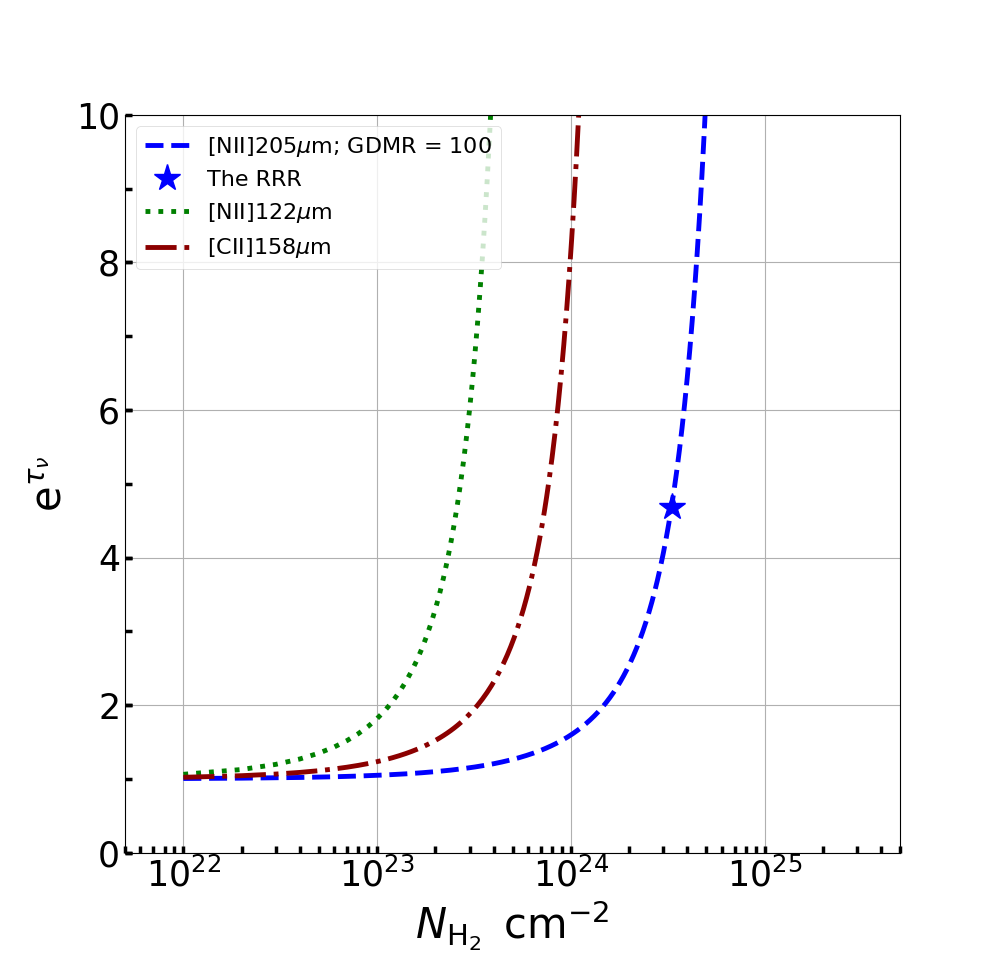}
    \caption{The dust attenuation correction as a function of mean molecular hydrogen column density. A single uniform dust screen approximation (with GDMR = 100; dust-emissivity spectral index, $\beta = 2.03$), evaluated at: the rest-wavelengths of the \Nplusa (green dotted line), the \Nplusb (blue dashed line) and [C\,{\sc ii}]158\mic (red dashed-dotted line) line emission, including the result for the \textit{RRR} (blue star).}
    \label{fig:dat}
\end{figure}

In the \textit{RRR} the dust opacity reaches unity at the rest-frame wavelength of $\lambda_{0} = $254 \micron. Such high opacity is consistent with that observed in other high-$z$ DSFGs \citep{Riechers2013Nature}, while slightly higher than e.g., AzTEC-3, which is a more normal star-forming galaxy at $z > 5$, \citep{riechers14}. The dust attenuation of the \Nplusa emission line is therefore not negligible, as pointed out by other studies of far-IR FSLs in quasars (QSOs) and DSFGs \citep{uzgil16, lamarche17,lamarche18}. We correct the line luminosity based on a single, uniform dust screen approximation, i.e., $L_{\rm [NII]205_{\rm un-att.}}\,=\, e^{\tau_{\rm 205\mu m}}\times L_{\rm [NII]205, obs.}$. From our best-fit SED model, the opacity at rest-frame 205\mic \, is $\tau_{\rm 205\mu m} = ( \lambda_{\rm 0}/\lambda)^{\beta} = 1.54$, which results in a uniform screen dust attenuation correction factor of ${\rm e}^{\tau_{\rm 205\mu m} } \simeq 4.67$. The following corrections are considered an upper limit along the line of sight. In the scenario where the emitting gas is well mixed with dust, the mixed gas/dust attenuation correction factor would be $ \tau_{\rm 205\mu m}/(1 - e^{- \tau_{\rm 205\mu m}} )$ = 1.96 .

Corrections to the observed far-IR FSL detections of the $z > 4$ systems are limited by the sparse sampling of their peak dust SED to accurately constrain the dust opacity at the relevant wavelength. The \textit{RRR}, with its well-sampled dust SED, allows us to constrain the mean hydrogen column density $N_{\rm H_{\rm 2}}$. In the following analysis, we assume both a fixed gas-to-dust-mass ratio, GDMR = 100, and a simple uniform dust screen.  The dust opacity is expressed in terms of the dust column density, $N_{\rm d}$, and $\kappa_{\rm \nu}$, the mass absorption coefficient \citep{weiss08}. We express $N_{\rm d}$ as $N_{\rm H_{\rm 2}}$ multiplied by the mass of molecular hydrogen, $m_{\rm H_{\rm 2}}$, divided by the GDMR:

\begin{equation}
\tau_{\nu} = \kappa_{\rm \nu} \times {\rm N_{d} = \kappa_{\rm \nu} \times \frac{N_{H_{2}} \times m_{\rm H_{\rm 2}}}{GDMR} },
\end{equation}
where
\begin{equation}
\kappa_{\nu} = 0.04\, (\nu/250 {\rm GHz})^{\beta}.
\end{equation}

Figure~\ref{fig:dat} plots the dust attenuation correction as a function of the mean molecular hydrogen column density N$_{\rm H_{\rm 2}}$ at rest-frame wavelengths corresponding to the \Nplusa, \Nplusb and [C\,{\sc ii}]158\mic emission lines. The equivalent \Htwo \, gas column density in the \textit{RRR} is $N_{\rm H_{\rm 2}} =\, 3.3 \times 10^{24} \, {\rm cm}^{-2}$.  

The \textit{RRR}, having high molecular gas column densities, will have corrections that can severely impact the use of both \Nplusa and \Nplusb emission lines as an electron density indicator. For example, based on this simple uniform screen approximation, the observed line ratio of \Nplusb / \Nplusa in the \textit{RRR} would need to be corrected by a factor of $e^{\rm \tau_{122\mu m}} / e^{\rm \tau_{205\mu m}} \approx 18$. An intrinsic \Nplusb / \Nplusa line ratio of $\sim 4-5$, corresponding to an electron density of $n_{\rm e^{-}} \simeq 200 $\, cm$^{-3}$ \citep[e.g. in the local starburst, M82; ][]{petuchowski94}, would thus yield an observed \Nplusb / \Nplusa value of (4-5)/18 = 0.3. Neglecting dust opacity, one would associate such low observed line ratio to un-physically low densities, as it would lie below the minimum theoretical line ratio of $\approx 0.6$, as derived for warm ionised regions with $n_{\rm e^{-}} << n_{\rm crit, 205\mu m}$ \citep{goldsmith15,hc16}. Naturally, the \textit{RRR}, but also all high-$z$ systems resembling the \textit{RRR}, i.e., having high column densities, would suffer from this effect. Dust opacity should not be neglected while studying far-IR FSL emission in high-$z$ DSFGs.

\subsubsection{Relative Cooling by [NII] 205$\mu$m Line Luminosity}

Using the \Nplusa and IR luminosities, we calculate the attenuation corrected $L_{\rm [NII]205\mu m} / L_{\rm IR} = 2.7 \pm 1.0 \times 10^{-4}$, assuming the same magnification factor for both luminosities. The vast majority of the local and high-\z galaxies do not correct for dust attenuation, therefore we use the apparent, attenuated value, $L_{\rm [NII]205\mu m} / L_{\rm IR} = 5.8 \pm 2.1 \times 10^{-5}$.  

\begin{figure}

	\includegraphics[width= \linewidth]{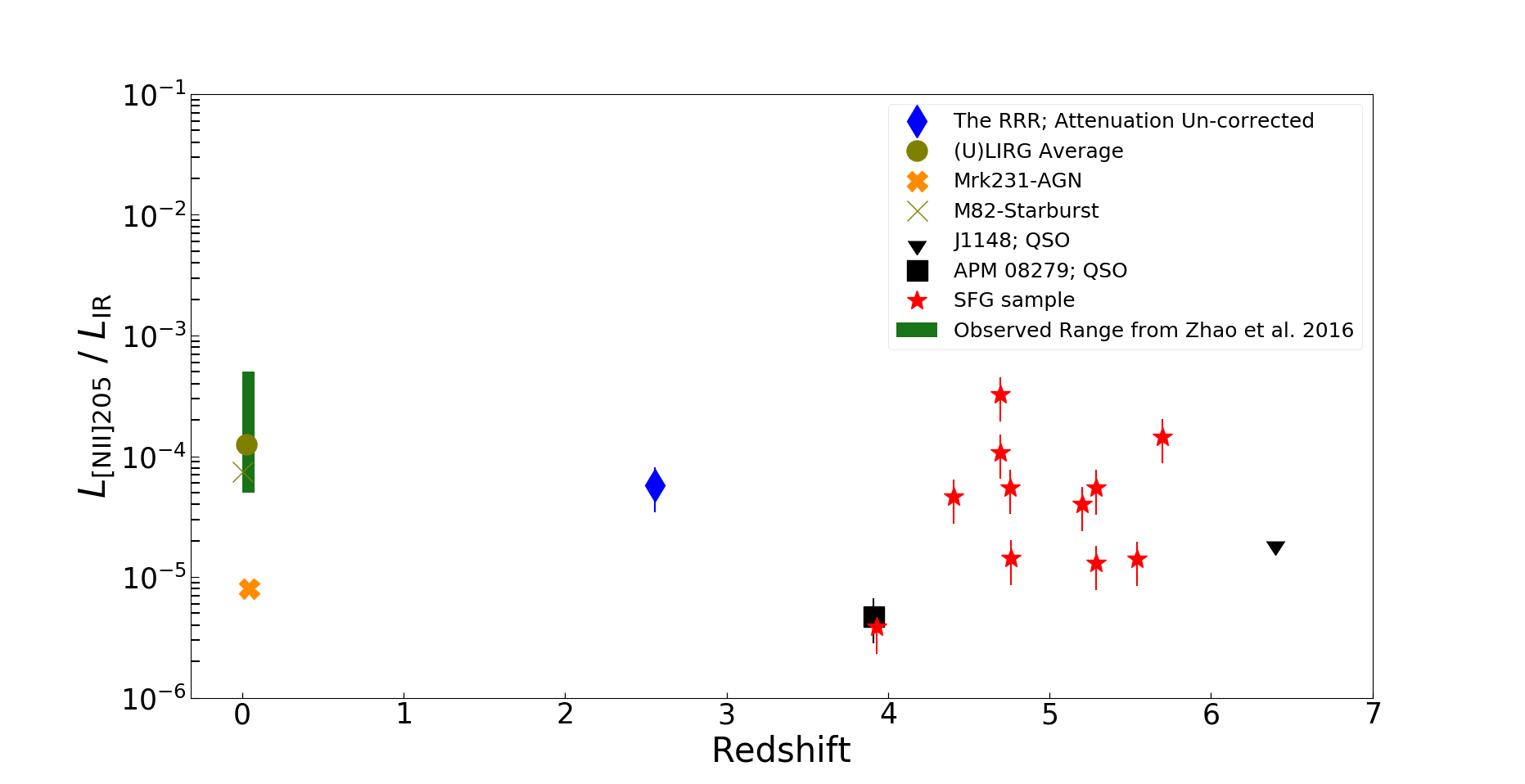}
    \caption{\Nplusa line luminosity to total IR (8-1000\mic) luminosity ratio in various samples, probing a broad redshift range: local starburst, M82 (green `x') and AGN, Mrk231 (orange `x'),  \citet{walter09,decarli12,decarli14,combes12,nagao12,bethermin16,pavesi16,pavesi18c,lu18} (red stars), including the attenuation corrected value for the \textit{RRR}. We show the predicted dispersion observed in the local Universe by \citet{zhao16} within a sample of galaxies with similar far-IR colour as the \textit{RRR} (green line).  }
    \label{fig:lumratvsz}
\end{figure}

As seen in Fig. \ref{fig:lumratvsz}, this attenuated  $L_{\rm [NII]205\mu m} / L_{IR}$ ratio for the \textit{RRR} is at the lower boundary of the mean range observed in local ULIRGs within the large scatter of $10^{-5} - 10^{-3}$ \citep{zhao16}. The large dispersion in the $L_{\rm [NII]205\mu m} / L_{\rm IR}$ ratio remains constant across all redshifts. Galaxies in Fig.~\ref{fig:lumratvsz} with the lowest values of $L_{\rm [NII]205\mu m} / L_{\rm IR}$ include strong QSOs at $z \sim 4$ \citep[e.g.][]{decarli12}, as well as the local AGN, MrK231, which has at least a 20\% AGN fraction contributing to its $L_{\rm IR}\approx 10^{12}$ L$_{\rm \odot}$ \citep{fischer10,dietrich18}. SFGs at high-$z$ have slightly large scatter, probing a range of up to a factor of five between low metallicity galaxies \citep{pavesi18b}, DSFGs and a Lyman-$\alpha$ Emitter \citep{decarli14}. 

The $L_{\rm [NII]205\mu m} / L_{IR}$ ratio of the \textit{RRR} more closely resembles that of local/high-z SB rather than that of local/high-$z$ QSO/AGNs. To first order, the global ISM within the \textit{RRR} is mostly powered by SF. We note, however, that this ratio is subject to a few caveats. Robust comparisons of this ratio between the \textit{RRR} and to other systems can be affected by individual variations in attenuation effects and hard ionising radiation fields that determine the relative \Nplusa line emission. The bolometric input to the total IR luminosity from a supermassive black hole accretion/activity could contaminate the apparent IR luminosities and reduce the observed line to total FIR luminosity, as seen in QSO-selected systems. However, this ratio may not decrease significantly if there is a narrow emission line region of an AGN contributing to the total \Nplusa line luminosity, as seen in the local selection of AGN via the [NII] 6584 $\si{\angstrom}$ / 6548 $\si{\angstrom}$ excess \citep{bpt}. Our dust SED model is consistent with that of a SB galaxy \citep{casey12, Magnelli2014}, yet the resemblance of such an SED can also be due to a large dust screen strongly attenuating the emission from an obscured, dusty AGN torus \citep[$T_{\rm d} \approx 500 $ K, Leung et al. 2018, submitted; ][]{siebenmorgen04,siebenmorgen15,feltre12,kirkpatrick17}. Thus, the spatially unresolved measurement of $L_{\rm [NII]205\mu m} / L_{IR}$ ratio cannot exclusively select an AGN from a SFG. 

In the local universe, the observed scatter correlates with the rest-frame log$(f_{\rm 70\mu m} /f_{\rm 160\mu m} )$ colour \citep{zhao13,zhao16}. SFGs with colder colours have an average of $L_{\rm [NII]205\mu m} / L_{\rm IR} \sim 3\times10^{-4}$, while star-forming/SB galaxies with warmer colours have average values of the $L_{\rm [NII]205\mu m} / L_{\rm IR}$ ratio of $\sim 5\times10^{-5}$ \citep{zhao16}. We show in Fig.~\ref{fig:lumratvsz} the range observed in the local Universe within galaxies with a similar FIR colour as the \textit{RRR} \citep[i.e. log$(f_{\rm 70\mu m} /f_{\rm 160\mu m} ) \sim 1.35$; see Fig. 3 in][]{zhao16}. The FIR colours can be interpreted as a proxy for the dust temperature. The ratio of the FIR FSL luminosity to IR luminosity in local ULIRGs, both with and without an AGN, reveals a so-called line-to-FIR-continuum ``deficit'', where the relative cooling efficiency of the line luminosity decreases with respect to the FIR continuum. This ``deficit'' increases for higher values of IR luminosity (warmer FIR colour) for \Nplusa, \Nplusb, [CII] 158$\mu$m, [OI] 63$\mu$m and [OIII] 88$\mu$m, with a two order of magnitude scatter for the \Nplus lines \citep{malhotra01, graciacarpio11, diazsantos17}. The nature of this deficit can strongly depend on the location of dust grains along the line of sight to the line emitting region \citep{diazsantos13,diazsantos17}, and that the ``deficit'' has a tight correlation to the relative compactness of the IR luminosity surface densities. This reflects the spatial concentration of dust-reprocessed far-UV through IR continuum photons, which diminishes the relative cooling power of the FIR FSL. Using the maximum radius of the \textit{RRR} in the reconstructed source-plane from the best-fit lens model of \citet[][$R_{\rm max}\simeq 2.6$ kpc]{geach18}, we infer a mean SFR surface density is $\Sigma_{\rm SFR_{IR}} = \frac{SFR_{\rm IR}}{\pi \, R_{\rm max}^{2}}$ of $\approx 120\, {\rm M_\odot yr^{-1} kpc^{-2}}$. The high SFR surface density may indicate why the apparent, attenuated ratio of $L_{\rm [NII]205\mu m} / L_{\rm IR}$ has a low value compared to the mean dispersion of local SFGs with similar rest-FIR colours (see green bar in Fig.~\ref{fig:lumratvsz}).

\subsection{CO Spectral Line Energy Distribution}
\begin{figure}
	\includegraphics[width=0.49\textwidth]{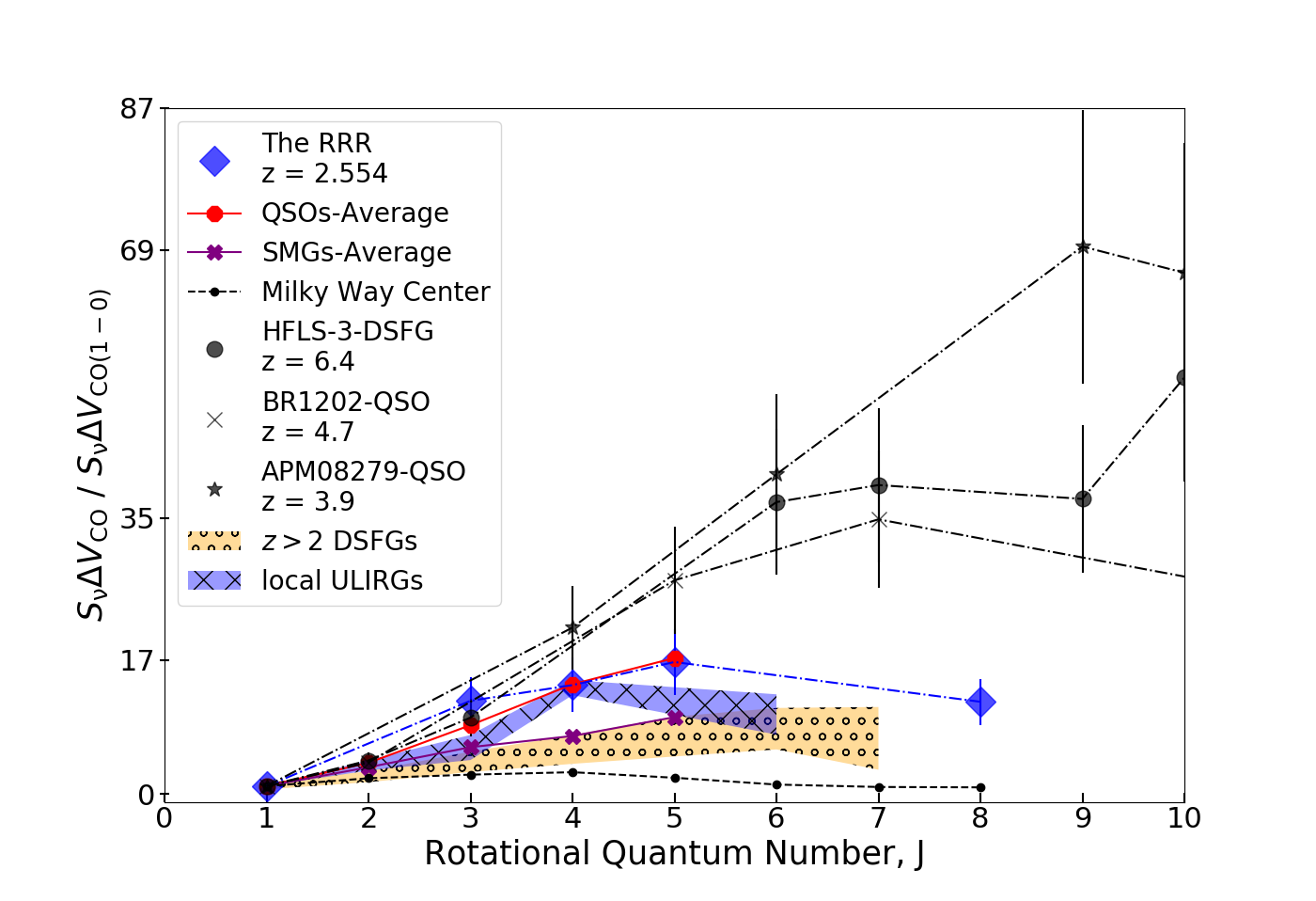}
    \caption{CO spectral line energy distribution (SLED), normalized to the ground-state CO(1-0) integrated flux density, of the \textit{RRR} (blue diamonds), including local ULIRGs \citep{pap12}, the Milky Way Centre \citep{fixsen99} and high-$z$ QSOs and DSFGs \citep{weiss07,salome12,bothwell13,Riechers2013Nature, carilli13}. }
    \label{fig:cosed}
\end{figure}

The spectral line energy distribution (SLED) of CO can be a tool to distinguish extreme, highly excited QSO galaxies from galaxies that have molecular gas excitation dominated by SF activity \citep{daddi15,carilli13}. Fig. \ref{fig:cosed} compares the CO(1-0) normalised SLED of the \textit{RRR} with the average spread amongst local ULIRGs, average DSFGs/SMGs, the Milky Way Galactic Centre, and well-known QSO powered systems at high-\z \citep{weiss07,fixsen99,pap12,bothwell13,riechers13}. Compared with the Milky Way Centre and the average dusty SFG, the \textit{RRR} shows high CO excitation. The CO SLED bears a resemblance to the even more extreme gas excitation in local ULIRGs \citep{pap12,mashian15,rosenberg15}, but not as high as the local starbursts, M82 \citep[][]{panuzzo10} and NGC 253 \citep{hd08}, or the more normal SFG, NGC 891--all of which peak at CO(7-6) \citep{nikola11}. The SLED is comparable to the average value of QSOs reported in the review by \citet{carilli13} out to J$_{\rm up} = 5$. Its molecular gas excitation hints at the existence of a strong heating source. To account for the strong mid-J CO lines observed in NGC 253 and NGC 891, both \citet[][]{hd08,nikola11} both invoke the need for strong mechanical heating/shocks from a turbulent star-forming environment \citep[see also][]{kamenetzky16,lu17}. The high-J turnover (at J$_{\rm up} \ge 5$) and tail of the CO SLED indicates that the CO excitation is, however, not as extreme as in the highly excited QSO systems \citep{weiss07,salome12}.

\subsection{Ionised and Molecular Gas Mass}
Local \Nplus measurements find that the electron density in both the Milky Way \citep[$n_{\rm e^{-}} = 33 \,{\rm cm^{-3}} $;][]{goldsmith15} and local (U)LIRGs \citep[$n_{\rm e^{-}} = 22 \, {\rm cm^{-3}} $; ][]{zhao16} is less than $ \sim 300 \,{\rm cm^{-3}}$. The minimum mass of ionised hydrogen can be approximated, after correcting for the line attenuation, using the high-temperature/high-density limit \citep[see][]{ferkinhoff11,decarli12}. Using Eq. 1 in \citet{decarli12}, we find $M_{\rm min}{\rm( H^{+}}) = 0.93 \pm 0.19 \times 10^{10} {\rm M_{\odot}}$. This assumes a gas phase nitrogen abundance of, $\chi (N) = N/ H= 9.3 \times 10^{-5}$ \citep{savage96}, determined by UV-absorption sight-lines towards massive stars in the Milky Way. We also assume that all the nitrogen is in the singly ionised state, i.e., $\chi (N)=\chi (N^{+})$. To calculate the relevant fraction of ionised to molecular gas mass, we use the measured $L'_{\rm CO(1-0)}$ line luminosity converted to a total molecular gas mass. Here we assume a CO line to molecular hydrogen gas mass conversion factor, $\alpha_{\rm CO} = 1$ M$_{\rm \odot}$ (K \kms pc$^{2}$)$^{-1}$, appropriate for ULIRGs \citep{Solomon2005,downes98,sanders96}\footnote{A range has been reported for individual ULIRGs: $\alpha = $0.6-2.6 \citep{downes98}}, and find $M_{\rm mol} = 3.67 \pm 0.9 \times 10^{10}$M$_{\rm \odot}$. To account for the total molecular gas mass, we correct by the additional mass-weighted contribution by He (1.36 $\times$ M$_{\rm H_{\rm 2}}$). The ionised to molecular mass fraction is 25\%, consistent with other actively star-forming high-$z$ systems \citep{ferkinhoff11, zhang18}. 

\section{Discussion}
\subsection{Ionised Nitrogen as a SFR tracer}

Despite the advantage of directly probing the ionising stars on timescales of $\sim$10 Myr, H$\alpha$ SFRs are often plagued by uncertainties from dust attenuation \citep{kennicutt98, calzetti07,Calzetti2000}. In dust-obscured galaxies, the attenuation-corrected H$\alpha$ luminosity measurements significantly underestimate the overall SFR \citep[e.g.][]{whit17} derived from the total IR (8-1000\mic) luminosity \citep{kennicutt98,kennicutt12}. In these dust-rich galaxies, the total IR luminosity has thus been seen as the ideal tracer of SFR because the dust-absorption cross section peaks at the wavelengths emitted by young stellar populations, and is re-emitted by dust in the far-IR wavelength regime \citep[e.g 40-120\mic, ][]{Helou1985}. IR-derived SFRs trace the \textit{characteristic, rather than instantaneous,} rates of SF on the order of 100 Myr, depending on the SF history. The total IR luminosity will be mostly dominated by the OB stellar population in starburst systems with less than a few 100 Myr gas consumption timescales \citep{kennicutt12}.

The far-IR FSLs have longer rest-wavelengths as compared to optical or near-IR tracers of SF, and the line photons may thus escape dust-obscured \Hplus regions without being absorbed. This motivates the use of far-IR FSLs as faithful tracers of the most recent SF \citep{fernont16}. For example, the \Nplus derived SFR probes \textit{quasi-instantaneous} star-formation rates corresponding to $\sim$ 30 Myr, i.e. the lifetimes of early B-stars. 

Locally, however, \Nplus may not be the ideal SFR tracer for galaxy integrated measurements, as it can trace a mixture of both ambient, diffuse ionised gas and ionised gas closely associated with SF. In contrast, \Nplus may be a more accurate tracer of the global SFR in gas-rich, SFGs at high-\z, as the latter contains star-forming environments that pervade the entire ISM. Indeed, these systems undergo rapid stellar mass assembly, with SF taking place throughout the entire galaxy \citep{magdis16,rujo16,elbaz18,ccc17}.

As first presented in \citet{ferkinhoff15}, there is a physically motivated relation between the SFR and \Nplus line emission for star-forming galaxies. This relation is further substantiated by an empirical relation between the observed \Nplus line emission and the IR-derived SFR \citep{zhao16}. In ionisation bounded, low-density \Hplus regions, the \Nplus line emission is proportional to the ionising photon rate (modulo the N/H ratio), which in turn is proportional to SF. The low-\z relation of \Nplus line luminosity to SFR can be extended out to high-\z only by making strong assumptions of both the fractional nitrogen abundance and ionised gas densities \citep{hc16}. Here we focus on the SFR estimate where the densities are below the critical density for \Nplusa emission line  \citep[44 \percc][]{goldsmith15}, and use Eq. 10 of \citet{hc16}. We assume the collisional excitation coefficients from \citet{tayal11}, which yields

\begin{equation}
 SFR_{\rm[N\,II]205\mu m} \,  [{\rm M_\odot \, yr^{-1}}] = 1.98 \times 10^{-7} \frac{ (N/H)_{\odot}}{(N^{+}/H^{+})} \frac{L_{\rm [N\,II]205\mu m}}{\rm L_{\odot}} .
\label{eq:SFRNII}
\end{equation}
We estimate an attenuation, and magnification, corrected SFR$_{\rm [NII]205\mu m}$ = 621 M$_{\rm \odot}$ yr$^{-1}$. However, both the nitrogen abundances in the \textit{RRR} are unknown, and will affect this measurement significantly. The \Nplusa derived SFR is about four times smaller than the traditional IR-derived SFR (Tab. \ref{table1}).

The SFR derived using Eq. \ref{eq:SFRNII} is, however, a lower-limit. The low-density assumption breaks down when the density reaches or even exceeds that of the critical density. This is a possibility for the \textit{RRR}, as a strong SB could result in increased electron densities and an overlap of \Hplus regions from widespread SF, yielding electron densities $> 10^{4}$\percc \citep[see \S~4.2.1 and theoretical \Nplus line ratio vs. $n_{\rm e^{-}}$ in][]{goldsmith15,hc16}. When the electron density is significantly higher than the critical density for the ground-state, the system is thermalised. Thus, the emission of photons is defined by the Boltzmann level population and the Einstein A-coefficient \footnote{We assume all of the nitrogen is within low-density \Hplus regions, as opposed to the hottest clusters of early O-type stars (which would result in most of the nitrogen residing in the \Nplusplus or [NIV], rather than \Nplus).}, saturating the \Nplusa line emission. Such high SFR surface density in the \textit{RRR} (see \S~4.2.2) implies the ISM has an electron density $n_{\rm e^{-}}\,>>\,100-1000$\percc \citep{hc16}, i.e. an order of magnitude higher than the critical density of the \Nplusa emission line. To confirm that the disagreement between these two SFR estimates is simply due to high electron density in the ISM of the \textit{RRR}, detection of the \Nplusb emission line will be needed.

\subsection{Co-Evolution of AGN/SF in the \textit{RRR}}
The known compact radio-mode AGN, inferred from high spatial resolution eMERLIN observations \citep{geach15}, does not seem to halt the intense SF activity of the \textit{RRR} \citep{geach18}. The unattenuated ratio of \Nplusa and IR luminosity is consistent with a SB galaxy, while the dust/CO SED also disfavors a strong AGN contaminating the IR spectrum \citep[e.g.][]{salome12,weiss07}. The similitude in line profile shapes of the multiple CO lines and the \Nplusa line points at the co-existence of warm ionised and cold molecular phases across kpc scales.  

As an example, the spatial co-existence of warm ionised regions traced by \Nplusa, and that of the dense and diffuse molecular gas traced by CO, can be considered similar to the gas phase mixing in the Central Molecular Zone (CMZ) on the scales of $100\,{\rm pc}$ \citep{kruijssen13,ginsburg18}, and even in the immediate vicinity of Sgr-A* on the scale of $1\,{\rm pc}$ \citep{moser17}. Hence, the Galactic centre region may serve as a high-$z$ galaxy analog \citep[see also][]{swinbank11,kruijssen13}. Thermal instability during AGN and SB phases in galaxies is one mechanism to explain how such a co-existence may be maintained over the long-term. \citet{rozanska14} and \citet{rozanska17} found that for certain parameter ranges (activity of a galactic nucleus, star cluster input), thermal instability operates\footnote{S-curve in the temperature--ionisation parameter plane ($T-\Sigma$), where the ionisation parameter $\Sigma$ is defined as $\Sigma=P_{\rm rad}/P_{\rm gas}$, where $P_{\rm rad}$ and $P_{\rm gas}$ are radiation and gas pressure, respectively.} and essentially leads to the formation of the two-phase (warm--cold) medium, which is rather stable given the long heating and cooling timescales \citep{field65}. This can be one of the main sources of cold gas formation under the presence of stellar, supernova and AGN feedbacks \citep{tenoriotagle13}. Our global measurements represent the galaxy integrated average of the CO and \Nplusa line emission, such that it is not possible to access the relevant physical scales to compare directly to the CMZ. \citet{geach15, geach18} identify a compact radio AGN ($<$250 pc) and a galactic disk traced by CO(4-3), extended over 2.5-3 kpc, therefore the CO(5-4) line emission is almost certainly associated with this molecular disk. However, the relative contributions of the AGN and the large-scale galactic disk to the observed \Nplusa line emission remains to be resolved. 

Such co-eval AGN/SF processes within galaxies is expected to be a part of the evolution of a massive galaxy such as the \textit{RRR}, depicted by a short-lived, merger-induced SB that catalyses high AGN activity and black-hole growth \citep[e.g.][]{hopkins08}. There is a range of co-eval AGN/SF processes that can be seen both locally and at high-$z$. In local systems, nuclear regions with high SF and low-AGN fractions are observed to co-exist based on various nebular line diagnostics e.g. \citep{dagostino18}, while there is an inferred quenching of SF in local AGN hosts residing in massive elliptical galaxies \citep{mcpartland19,baron18}. The increased excitation conditions within the narrow emission line regions of an unobscured AGN can ionise the entire ISM \citep{greene11}, and potentially quench SF. Therefore the use of ionised nitrogen as a tracer of SF may be unreliable because of the change in ionisation structure of nitrogen (depending on the slope of the ionisation parameter) in the presence of such a strong heating source. High-$z$ galaxies, however, with large reservoirs of molecular gas ($> 10^{9-10}$M$_{\rm \odot}$) can sustain ongoing SF even in the most extreme, optically bright, broad-line QSO systems \citep{az16,glikman15,Cresci2015}. A systematic study of $>$100 gravitationally lensed QSOs ($z\sim1-4$), \citet{stacey17} found most have both SF and AGN activity. Our conclusions in \S~4.2.2 and 4.3 suggest it is possible for both AGN and SB activity to co-exist, and this may be due to both thermal instability and the large molecular gas reservoir in the \textit{RRR}.

\section{Summary and Conclusions}
We present the detection of \Nplusa in a strongly lensed AGN/SB galaxy at $z = 2.5535 \pm 0.0006$, obtained using the \textit{APEX} telescope. We complement this detection with multiple CO line transitions (CO 1-0, 5-4, 8-7) to examine the global properties of the ionised and molecular gas in the \textit{RRR}. Our main conclusions are:\\

$\bullet \,$ The line profiles for the CO and the \Nplusa emission lines have similar velocity components that can be explained by shared volumes, i.e. molecular clouds well-mixed with \Hplus regions, suggesting the majority of the strong \Nplusa detection is associated with SF.\\

$\bullet \,$ The non-negligible dust attenuation at rest-frame 205\mic in the \textit{RRR} suggests that these corrections need to be accounted for when interpreting far-IR FSLs in dust-rich systems at high-z. Assuming a uniform dust screen approximation results in a dust attenuation correction, $ {\rm e}^{\tau_{205\mu m} }$, of $\sim 4.67$ for the \textit{RRR}. This implies a mean H$_{\rm 2}$ gas column density $> 10^{24}$\, cm$^{-2}$, assuming a molecular gas-to-dust mass ratio of 100. \\

$\bullet \,$ We derived an attenuation corrected, minimum ionised gas mass, M$_{\rm min}$(H$^{+}$)$ = 0.89 \times 10^{10} (\frac{15}{\mu})$ M$_{\odot}$, assuming a high-density / high-temperature limit. This ionised gas mass corresponds to about 25\% of the total molecular gas mass derived using the measured CO(1-0) line luminosity and $\alpha_{\rm CO} = 1$ M$_{\rm \odot}$ (K \kms pc$^{2}$)$^{-1}$. \\

$\bullet \,$ The attenuation corrected value of $L_{\rm [NII]205\mu m} / L_{\rm IR} = 2.7 \pm 1.0 \times 10^{-4}$, resembles the average values of SFGs rather than those with a known QSO influence. \\

$\bullet \,$ The IR SFR, SFR$_{\rm IR}$ =  2482$\pm$ 992 M$_{\rm \odot}$ yr$^{-1}$, is a factor of four larger than the lower-limit SFR estimate from the attenuation corrected, \Nplusa line luminosity in the low-density regime: SFR$_{\rm [NII]205\mu m} = $621 M$_{\rm \odot}$ yr$^{-1}$. This suggests the electron density is significantly high, or the assumed nitrogen abundance is significantly lower.  \\

Utilising the \Nplusa line as a tracer of SF has a strong physical motivation, although the reliable application of local relations requires extensive calibration for high-$z$ dusty SFGs. Future spatially resolved \Nplusa and \Nplusb observations would help to isolate low-density vs. high-density \Hplus complexes in the warm ionised medium \citep[as seen in ][]{spinoglio15,zhao16} in order to aid future interpretations in this system. 

\section*{Acknowledgements}
The authors would like to thank the referee for her/his comments and suggestions to enhance the quality of the manuscript. KCH extends his appreciation to the entire facility staff/observers/operators at the GBT, IRAM 30m and APEX for making the accommodation and observing welcoming and successful.The Green Bank Observatory is a facility of the National Science Foundation operated under cooperative agreement by Associated Universities, Inc. IRAM is supported by INSU/CNRS (France), MPG (Germany) and IGN (Spain). This work is carried out within the Collaborative Research Centre 956, sub-project [A1, C4], funded by the Deutsche Forschungsgemeinschaft (DFG). D.R. acknowledges support from the National Science Foundation under grant number AST-1614213.This publication is based on data acquired with the Atacama Pathfinder Experiment (APEX) Telescope. APEX is a collaboration between the Max-Planck-Institut fur Radioastronomie, the European Southern Observatory, and the Onsala Space Observatory. IRAM is supported by INSU/CNRS (France), MPG (Germany) and IGN (Spain).TKDL acknowledges support from the NSF through award SOSPA4-009 from 
the NRAO and support from the Simons Foundation. The Flatiron Institute 
is supported by the Simons Foundation.
 



\bibliographystyle{mnras}
\bibliography{bigbibs} 

\begin{thebibliography}{}
\makeatletter
\relax
\def\mn@urlcharsother{\let\do\@makeother \do\$\do\&\do\#\do\^\do\_\do\%\do\~}
\def\mn@doi{\begingroup\mn@urlcharsother \@ifnextchar [ {\mn@doi@}
  {\mn@doi@[]}}
\def\mn@doi@[#1]#2{\def\@tempa{#1}\ifx\@tempa\@empty \href
  {http://dx.doi.org/#2} {doi:#2}\else \href {http://dx.doi.org/#2} {#1}\fi
  \endgroup}
\def\mn@eprint#1#2{\mn@eprint@#1:#2::\@nil}
\def\mn@eprint@arXiv#1{\href {http://arxiv.org/abs/#1} {{\tt arXiv:#1}}}
\def\mn@eprint@dblp#1{\href {http://dblp.uni-trier.de/rec/bibtex/#1.xml}
  {dblp:#1}}
\def\mn@eprint@#1:#2:#3:#4\@nil{\def\@tempa {#1}\def\@tempb {#2}\def\@tempc
  {#3}\ifx \@tempc \@empty \let \@tempc \@tempb \let \@tempb \@tempa \fi \ifx
  \@tempb \@empty \def\@tempb {arXiv}\fi \@ifundefined
  {mn@eprint@\@tempb}{\@tempb:\@tempc}{\expandafter \expandafter \csname
  mn@eprint@\@tempb\endcsname \expandafter{\@tempc}}}

\bibitem[\protect\citeauthoryear{{Alaghband-Zadeh}, {Banerji}, {Hewett}  \&
  {McMahon}}{{Alaghband-Zadeh} et~al.}{2016}]{az16}
{Alaghband-Zadeh} S.,  {Banerji} M.,  {Hewett} P.~C.,   {McMahon} R.~G.,  2016,
  \mn@doi [\mnras] {10.1093/mnras/stw682}, \href
  {https://ui.adsabs.harvard.edu/#abs/2016MNRAS.459..999A} {459, 999}

\bibitem[\protect\citeauthoryear{{Baldwin}, {Phillips}  \&
  {Terlevich}}{{Baldwin} et~al.}{1981}]{bpt}
{Baldwin} J.~A.,  {Phillips} M.~M.,   {Terlevich} R.,  1981, \mn@doi [\pasp]
  {10.1086/130766}, \href
  {https://ui.adsabs.harvard.edu/abs/1981PASP...93....5B} {93, 5}

\bibitem[\protect\citeauthoryear{{Baron} et~al.,}{{Baron}
  et~al.}{2018}]{baron18}
{Baron} D.,  et~al., 2018, \mn@doi [\mnras] {10.1093/mnras/sty2113}, \href
  {https://ui.adsabs.harvard.edu/#abs/2018MNRAS.480.3993B} {480, 3993}

\bibitem[\protect\citeauthoryear{{Bennett} et~al.,}{{Bennett}
  et~al.}{1994}]{bennett94}
{Bennett} C.~L.,  et~al., 1994, \mn@doi [\apj] {10.1086/174761}, \href
  {https://ui.adsabs.harvard.edu/#abs/1994ApJ...434..587B} {434, 587}

\bibitem[\protect\citeauthoryear{{Bennett}, {Larson}, {Weiland}  \&
  {Hinshaw}}{{Bennett} et~al.}{2014}]{bennett14}
{Bennett} C.~L.,  {Larson} D.,  {Weiland} J.~L.,   {Hinshaw} G.,  2014, \mn@doi
  [\apj] {10.1088/0004-637X/794/2/135}, \href
  {http://adsabs.harvard.edu/abs/2014ApJ...794..135B} {794, 135}

\bibitem[\protect\citeauthoryear{{B{\'e}thermin} et~al.,}{{B{\'e}thermin}
  et~al.}{2016}]{bethermin16}
{B{\'e}thermin} M.,  et~al., 2016, \mn@doi [\aap]
  {10.1051/0004-6361/201527739}, \href
  {http://adsabs.harvard.edu/abs/2016A\%26A...586L...7B} {586, L7}

\bibitem[\protect\citeauthoryear{{Bothwell} et~al.,}{{Bothwell}
  et~al.}{2013}]{bothwell13}
{Bothwell} M.~S.,  et~al., 2013, \mn@doi [\mnras] {10.1093/mnras/sts562}, \href
  {http://adsabs.harvard.edu/abs/2013MNRAS.429.3047B} {429, 3047}

\bibitem[\protect\citeauthoryear{{Brisbin}, {Ferkinhoff}, {Nikola}, {Parshley},
  {Stacey}, {Spoon}, {Hailey-Dunsheath}  \& {Verma}}{{Brisbin}
  et~al.}{2015}]{brisbin15}
{Brisbin} D.,  {Ferkinhoff} C.,  {Nikola} T.,  {Parshley} S.,  {Stacey} G.~J.,
  {Spoon} H.,  {Hailey-Dunsheath} S.,   {Verma} A.,  2015, \mn@doi [\apj]
  {10.1088/0004-637X/799/1/13}, \href
  {https://ui.adsabs.harvard.edu/#abs/2015ApJ...799...13B} {799, 13}

\bibitem[\protect\citeauthoryear{{Brott} et~al.,}{{Brott}
  et~al.}{2011}]{brott11a}
{Brott} I.,  et~al., 2011, \mn@doi [\aap] {10.1051/0004-6361/201016113}, \href
  {https://ui.adsabs.harvard.edu/#abs/2011A&A...530A.115B} {530, A115}

\bibitem[\protect\citeauthoryear{{Ca{\~n}ameras} et~al.}{{Ca{\~n}ameras}
  et~al.}{2015}]{canameras15}
{Ca{\~n}ameras} R.,  et~al., 2015, \mn@doi [\aap]
  {10.1051/0004-6361/201425128}, \href
  {http://adsabs.harvard.edu/abs/2015A\%26A...581A.105C} {581, A105}

\bibitem[\protect\citeauthoryear{{Calzetti}, {Armus}, {Bohlin}, {Kinney},
  {Koornneef}  \& {Storchi-Bergmann}}{{Calzetti} et~al.}{2000}]{Calzetti2000}
{Calzetti} D.,  {Armus} L.,  {Bohlin} R.~C.,  {Kinney} A.~L.,  {Koornneef} J.,
   {Storchi-Bergmann} T.,  2000, \mn@doi [\apj] {10.1086/308692}, \href
  {http://adsabs.harvard.edu/abs/2000ApJ...533..682C} {533, 682}

\bibitem[\protect\citeauthoryear{{Calzetti} et~al.,}{{Calzetti}
  et~al.}{2007}]{calzetti07}
{Calzetti} D.,  et~al., 2007, \mn@doi [\apj] {10.1086/520082}, \href
  {https://ui.adsabs.harvard.edu/#abs/2007ApJ...666..870C} {666, 870}

\bibitem[\protect\citeauthoryear{{Carilli} \& {Walter}}{{Carilli} \&
  {Walter}}{2013}]{carilli13}
{Carilli} C.~L.,  {Walter} F.,  2013, \mn@doi [\araa]
  {10.1146/annurev-astro-082812-140953}, \href
  {http://adsabs.harvard.edu/abs/2013ARA\%26A..51..105C} {51, 105}

\bibitem[\protect\citeauthoryear{{Casey} et~al.}{{Casey}
  et~al.}{2012}]{casey12}
{Casey} C.~M.,  et~al., 2012, \mn@doi [\apj] {10.1088/0004-637X/761/2/139},
  \href {http://adsabs.harvard.edu/abs/2012ApJ...761..139C} {761, 139}

\bibitem[\protect\citeauthoryear{{Chen} et~al.,}{{Chen} et~al.}{2017}]{ccc17}
{Chen} C.-C.,  et~al., 2017, \mn@doi [\apj] {10.3847/1538-4357/aa863a}, \href
  {https://ui.adsabs.harvard.edu/#abs/2017ApJ...846..108C} {846, 108}

\bibitem[\protect\citeauthoryear{{Cicone} et~al.,}{{Cicone}
  et~al.}{2014}]{cicone14}
{Cicone} C.,  et~al., 2014, \mn@doi [\aap] {10.1051/0004-6361/201322464}, \href
  {https://ui.adsabs.harvard.edu/\#abs/2014A&A...562A..21C} {562, A21}

\bibitem[\protect\citeauthoryear{{Cicone} et~al.,}{{Cicone}
  et~al.}{2015}]{cicone15}
{Cicone} C.,  et~al., 2015, \mn@doi [\aap] {10.1051/0004-6361/201424980}, \href
  {https://ui.adsabs.harvard.edu/\#abs/2015A&A...574A..14C} {574, A14}

\bibitem[\protect\citeauthoryear{{Cicone} et~al.,}{{Cicone}
  et~al.}{2018}]{cicone18}
{Cicone} C.,  et~al., 2018, \mn@doi [\apj] {10.3847/1538-4357/aad32a}, \href
  {https://ui.adsabs.harvard.edu/#abs/2018ApJ...863..143C} {863, 143}

\bibitem[\protect\citeauthoryear{{Colgan}, {Haas}, {Erickson}, {Rubin},
  {Simpson}  \& {Russell}}{{Colgan} et~al.}{1993}]{colgan93}
{Colgan} S. W.~J.,  {Haas} M.~R.,  {Erickson} E.~F.,  {Rubin} R.~H.,  {Simpson}
  J.~P.,   {Russell} R.~W.,  1993, \mn@doi [\apj] {10.1086/172991}, \href
  {https://ui.adsabs.harvard.edu/#abs/1993ApJ...413..237C} {413, 237}

\bibitem[\protect\citeauthoryear{{Combes} et~al.,}{{Combes}
  et~al.}{2012}]{combes12}
{Combes} F.,  et~al., 2012, \mn@doi [\aap] {10.1051/0004-6361/201118750}, \href
  {http://adsabs.harvard.edu/abs/2012A\%26A...538L...4C} {538, L4}

\bibitem[\protect\citeauthoryear{{Cormier} et~al.,}{{Cormier}
  et~al.}{2015}]{cormier15}
{Cormier} D.,  et~al., 2015, \mn@doi [\aap] {10.1051/0004-6361/201425207},
  \href {https://ui.adsabs.harvard.edu/#abs/2015A&A...578A..53C} {578, A53}

\bibitem[\protect\citeauthoryear{{Cresci} et~al.,}{{Cresci}
  et~al.}{2015}]{Cresci2015}
{Cresci} G.,  et~al., 2015, \mn@doi [\apj] {10.1088/0004-637X/799/1/82}, \href
  {http://cdsads.u-strasbg.fr/abs/2015ApJ...799...82C} {799, 82}

\bibitem[\protect\citeauthoryear{{Crowther}}{{Crowther}}{2007}]{crowther07}
{Crowther} P.~A.,  2007, \mn@doi [Annual Review of Astronomy and Astrophysics]
  {10.1146/annurev.astro.45.051806.110615}, \href
  {https://ui.adsabs.harvard.edu/#abs/2007ARA&A..45..177C} {45, 177}

\bibitem[\protect\citeauthoryear{{D'Agostino}, {Poetrodjojo}, {Ho}, {Groves},
  {Kewley}, {Madore}, {Rich}  \& {Seibert}}{{D'Agostino}
  et~al.}{2018}]{dagostino18}
{D'Agostino} J.~J.,  {Poetrodjojo} H.,  {Ho} I.~T.,  {Groves} B.,  {Kewley} L.,
   {Madore} B.~F.,  {Rich} J.,   {Seibert} M.,  2018, \mn@doi [\mnras]
  {10.1093/mnras/sty1676}, \href
  {https://ui.adsabs.harvard.edu/#abs/2018MNRAS.479.4907D} {479, 4907}

\bibitem[\protect\citeauthoryear{{Daddi} et~al.,}{{Daddi}
  et~al.}{2015}]{daddi15}
{Daddi} E.,  et~al., 2015, \mn@doi [\aap] {10.1051/0004-6361/201425043}, \href
  {http://adsabs.harvard.edu/abs/2015A\%26A...577A..46D} {577, A46}

\bibitem[\protect\citeauthoryear{{Dale}, {Helou}, {Contursi}, {Silbermann}  \&
  {Kolhatkar}}{{Dale} et~al.}{2001}]{dale01}
{Dale} D.~A.,  {Helou} G.,  {Contursi} A.,  {Silbermann} N.~A.,   {Kolhatkar}
  S.,  2001, \mn@doi [\apj] {10.1086/319077}, \href
  {https://ui.adsabs.harvard.edu/\#abs/2001ApJ...549..215D} {549, 215}

\bibitem[\protect\citeauthoryear{{Dalla Vecchia} \& {Schaye}}{{Dalla Vecchia}
  \& {Schaye}}{2008}]{dallavecchia08}
{Dalla Vecchia} C.,  {Schaye} J.,  2008, \mn@doi [\mnras]
  {10.1111/j.1365-2966.2008.13322.x}, \href
  {https://ui.adsabs.harvard.edu/\#abs/2008MNRAS.387.1431D} {387, 1431}

\bibitem[\protect\citeauthoryear{{Decarli} et~al.,}{{Decarli}
  et~al.}{2012}]{decarli12}
{Decarli} R.,  et~al., 2012, \mn@doi [\apj] {10.1088/0004-637X/752/1/2}, \href
  {https://ui.adsabs.harvard.edu/#abs/2012ApJ...752....2D} {752, 2}

\bibitem[\protect\citeauthoryear{{Decarli} et~al.,}{{Decarli}
  et~al.}{2014}]{decarli14}
{Decarli} R.,  et~al., 2014, \mn@doi [\apj] {10.1088/2041-8205/782/2/L17},
  \href {https://ui.adsabs.harvard.edu/#abs/2014ApJ...782L..17D} {782, L17}

\bibitem[\protect\citeauthoryear{{D{\'\i}az-Santos} et~al.,}{{D{\'\i}az-Santos}
  et~al.}{2013}]{diazsantos13}
{D{\'\i}az-Santos} T.,  et~al., 2013, \mn@doi [\apj]
  {10.1088/0004-637X/774/1/68}, \href
  {https://ui.adsabs.harvard.edu/#abs/2013ApJ...774...68D} {774, 68}

\bibitem[\protect\citeauthoryear{{D{\'\i}az-Santos} et~al.,}{{D{\'\i}az-Santos}
  et~al.}{2017}]{diazsantos17}
{D{\'\i}az-Santos} T.,  et~al., 2017, \mn@doi [\apj]
  {10.3847/1538-4357/aa81d7}, \href
  {https://ui.adsabs.harvard.edu/#abs/2017ApJ...846...32D} {846, 32}

\bibitem[\protect\citeauthoryear{{Dietrich} et~al.,}{{Dietrich}
  et~al.}{2018}]{dietrich18}
{Dietrich} J.,  et~al., 2018, \mn@doi [\mnras] {10.1093/mnras/sty2056}, \href
  {https://ui.adsabs.harvard.edu/#abs/2018MNRAS.480.3562D} {480, 3562}

\bibitem[\protect\citeauthoryear{{Downes} \& {Solomon}}{{Downes} \&
  {Solomon}}{1998}]{downes98}
{Downes} D.,  {Solomon} P.~M.,  1998, \mn@doi [\apj] {10.1086/306339}, \href
  {https://ui.adsabs.harvard.edu/#abs/1998ApJ...507..615D} {507, 615}

\bibitem[\protect\citeauthoryear{{Ekstr{\"o}m} et~al.,}{{Ekstr{\"o}m}
  et~al.}{2012}]{ekstrom12}
{Ekstr{\"o}m} S.,  et~al., 2012, \mn@doi [\aap] {10.1051/0004-6361/201117751},
  \href {http://adsabs.harvard.edu/abs/2012A%26A...537A.146E} {537, A146}

\bibitem[\protect\citeauthoryear{{Elbaz} et~al.,}{{Elbaz}
  et~al.}{2018}]{elbaz18}
{Elbaz} D.,  et~al., 2018, \mn@doi [\aap] {10.1051/0004-6361/201732370}, \href
  {https://ui.adsabs.harvard.edu/#abs/2018A&A...616A.110E} {616, A110}

\bibitem[\protect\citeauthoryear{{Farrah} et~al.,}{{Farrah}
  et~al.}{2013}]{farrah13}
{Farrah} D.,  et~al., 2013, \mn@doi [\apj] {10.1088/0004-637X/776/1/38}, \href
  {https://ui.adsabs.harvard.edu/#abs/2013ApJ...776...38F} {776, 38}

\bibitem[\protect\citeauthoryear{{Feltre}, {Hatziminaoglou}, {Fritz}  \&
  {Franceschini}}{{Feltre} et~al.}{2012}]{feltre12}
{Feltre} A.,  {Hatziminaoglou} E.,  {Fritz} J.,   {Franceschini} A.,  2012,
  \mn@doi [\mnras] {10.1111/j.1365-2966.2012.21695.x}, \href
  {https://ui.adsabs.harvard.edu/#abs/2012MNRAS.426..120F} {426, 120}

\bibitem[\protect\citeauthoryear{{Ferkinhoff}, {Hailey-Dunsheath}, {Nikola},
  {Parshley}, {Stacey}, {Benford}  \& {Staguhn}}{{Ferkinhoff}
  et~al.}{2010}]{ferkinhoff10}
{Ferkinhoff} C.,  {Hailey-Dunsheath} S.,  {Nikola} T.,  {Parshley} S.~C.,
  {Stacey} G.~J.,  {Benford} D.~J.,   {Staguhn} J.~G.,  2010, \mn@doi [\apj]
  {10.1088/2041-8205/714/1/L147}, \href
  {https://ui.adsabs.harvard.edu/#abs/2010ApJ...714L.147F} {714, L147}

\bibitem[\protect\citeauthoryear{{Ferkinhoff} et~al.,}{{Ferkinhoff}
  et~al.}{2011}]{ferkinhoff11}
{Ferkinhoff} C.,  et~al., 2011, \mn@doi [\apj] {10.1088/2041-8205/740/1/L29},
  \href {https://ui.adsabs.harvard.edu/#abs/2011ApJ...740L..29F} {740, L29}

\bibitem[\protect\citeauthoryear{{Ferkinhoff}, {Brisbin}, {Nikola}, {Stacey},
  {Sheth}, {Hailey-Dunsheath}  \& {Falgarone}}{{Ferkinhoff}
  et~al.}{2015}]{ferkinhoff15}
{Ferkinhoff} C.,  {Brisbin} D.,  {Nikola} T.,  {Stacey} G.~J.,  {Sheth} K.,
  {Hailey-Dunsheath} S.,   {Falgarone} E.,  2015, \mn@doi [\apj]
  {10.1088/0004-637X/806/2/260}, \href
  {https://ui.adsabs.harvard.edu/#abs/2015ApJ...806..260F} {806, 260}

\bibitem[\protect\citeauthoryear{{Fern{\'a}ndez-Ontiveros}, {Spinoglio},
  {Pereira-Santaella}, {Malkan}, {Andreani}  \&
  {Dasyra}}{{Fern{\'a}ndez-Ontiveros} et~al.}{2016}]{fernont16}
{Fern{\'a}ndez-Ontiveros} J.~A.,  {Spinoglio} L.,  {Pereira-Santaella} M.,
  {Malkan} M.~A.,  {Andreani} P.,   {Dasyra} K.~M.,  2016, \mn@doi [The
  Astrophysical Journal Supplement Series] {10.3847/0067-0049/226/2/19}, \href
  {https://ui.adsabs.harvard.edu/#abs/2016ApJS..226...19F} {226, 19}

\bibitem[\protect\citeauthoryear{{Field}}{{Field}}{1965}]{field65}
{Field} G.~B.,  1965, \mn@doi [\apj] {10.1086/148317}, \href
  {http://adsabs.harvard.edu/abs/1965ApJ...142..531F} {142, 531}

\bibitem[\protect\citeauthoryear{{Fischer} et~al.,}{{Fischer}
  et~al.}{2010}]{fischer10}
{Fischer} J.,  et~al., 2010, \mn@doi [\aap] {10.1051/0004-6361/201014676},
  \href {https://ui.adsabs.harvard.edu/#abs/2010A&A...518L..41F} {518, L41}

\bibitem[\protect\citeauthoryear{{Fixsen}, {Bennett}  \& {Mather}}{{Fixsen}
  et~al.}{1999}]{fixsen99}
{Fixsen} D.~J.,  {Bennett} C.~L.,   {Mather} J.~C.,  1999, \mn@doi [\apj]
  {10.1086/307962}, \href {http://adsabs.harvard.edu/abs/1999ApJ...526..207F}
  {526, 207}

\bibitem[\protect\citeauthoryear{{Frayer}, {Maddalena}, {Ivison}, {Smail},
  {Blain}  \& {Vanden Bout}}{{Frayer} et~al.}{2018}]{frayer18}
{Frayer} D.~T.,  {Maddalena} R.~J.,  {Ivison} R.~J.,  {Smail} I.,  {Blain}
  A.~W.,   {Vanden Bout} P.,  2018, \mn@doi [\apj] {10.3847/1538-4357/aac49a},
  \href {https://ui.adsabs.harvard.edu/#abs/2018ApJ...860...87F} {860, 87}

\bibitem[\protect\citeauthoryear{{Fu}, {Jullo}, {Cooray}, {Bussmann}, {Ivison}
  et~al.}{{Fu} et~al.}{2012}]{fu12}
{Fu} H.,  {Jullo} E.,  {Cooray} A.,  {Bussmann} R.~S.,  {Ivison} R.~J.,
  et~al., 2012, \mn@doi [\apj] {10.1088/0004-637X/753/2/134}, \href
  {http://adsabs.harvard.edu/abs/2012ApJ...753..134F} {753, 134}

\bibitem[\protect\citeauthoryear{{Geach} et~al.}{{Geach}
  et~al.}{2015}]{geach15}
{Geach} J.~E.,  et~al., 2015, \mn@doi [\mnras] {10.1093/mnras/stv1243}, \href
  {http://adsabs.harvard.edu/abs/2015MNRAS.452..502G} {452, 502}

\bibitem[\protect\citeauthoryear{{Geach}, {Ivison}, {Dye}  \& {Oteo}}{{Geach}
  et~al.}{2018}]{geach18}
{Geach} J.~E.,  {Ivison} R.~J.,  {Dye} S.,   {Oteo} I.,  2018, \mn@doi [\apj]
  {10.3847/2041-8213/aae375}, \href
  {https://ui.adsabs.harvard.edu/#abs/2018ApJ...866L..12G} {866, L12}

\bibitem[\protect\citeauthoryear{{Genzel} et~al.,}{{Genzel}
  et~al.}{2010}]{genzel10}
{Genzel} R.,  et~al., 2010, \mn@doi [\mnras]
  {10.1111/j.1365-2966.2010.16969.x}, \href
  {http://adsabs.harvard.edu/abs/2010MNRAS.407.2091G} {407, 2091}

\bibitem[\protect\citeauthoryear{{Ginsburg} et~al.,}{{Ginsburg}
  et~al.}{2018}]{ginsburg18}
{Ginsburg} A.,  et~al., 2018, \mn@doi [\apj] {10.3847/1538-4357/aaa6d4}, \href
  {http://adsabs.harvard.edu/abs/2018ApJ...853..171G} {853, 171}

\bibitem[\protect\citeauthoryear{{Glikman}, {Simmons}, {Mailly}, {Schawinski},
  {Urry}  \& {Lacy}}{{Glikman} et~al.}{2015}]{glikman15}
{Glikman} E.,  {Simmons} B.,  {Mailly} M.,  {Schawinski} K.,  {Urry} C.~M.,
  {Lacy} M.,  2015, \mn@doi [\apj] {10.1088/0004-637X/806/2/218}, \href
  {https://ui.adsabs.harvard.edu/#abs/2015ApJ...806..218G} {806, 218}

\bibitem[\protect\citeauthoryear{{Goldsmith}, {Y{\i}ld{\i}z}, {Langer}  \&
  {Pineda}}{{Goldsmith} et~al.}{2015}]{goldsmith15}
{Goldsmith} P.~F.,  {Y{\i}ld{\i}z} U.~A.,  {Langer} W.~D.,   {Pineda} J.~L.,
  2015, \mn@doi [\apj] {10.1088/0004-637X/814/2/133}, \href
  {https://ui.adsabs.harvard.edu/#abs/2015ApJ...814..133G} {814, 133}

\bibitem[\protect\citeauthoryear{{Graci{\'a}-Carpio}
  et~al.,}{{Graci{\'a}-Carpio} et~al.}{2011}]{graciacarpio11}
{Graci{\'a}-Carpio} J.,  et~al., 2011, \mn@doi [\apj]
  {10.1088/2041-8205/728/1/L7}, \href
  {https://ui.adsabs.harvard.edu/#abs/2011ApJ...728L...7G} {728, L7}

\bibitem[\protect\citeauthoryear{{Greene}, {Zakamska}, {Ho}  \&
  {Barth}}{{Greene} et~al.}{2011}]{greene11}
{Greene} J.~E.,  {Zakamska} N.~L.,  {Ho} L.~C.,   {Barth} A.~J.,  2011, \mn@doi
  [\apj] {10.1088/0004-637X/732/1/9}, \href
  {https://ui.adsabs.harvard.edu/#abs/2011ApJ...732....9G} {732, 9}

\bibitem[\protect\citeauthoryear{{G{\"u}sten}, {Nyman}, {Schilke}, {Menten},
  {Cesarsky}  \& {Booth}}{{G{\"u}sten} et~al.}{2006}]{guesten06}
{G{\"u}sten} R.,  {Nyman} L.~{\r{A}}.,  {Schilke} P.,  {Menten} K.,  {Cesarsky}
  C.,   {Booth} R.,  2006, \mn@doi [\aap] {10.1051/0004-6361:20065420}, \href
  {https://ui.adsabs.harvard.edu/#abs/2006A&A...454L..13G} {454, L13}

\bibitem[\protect\citeauthoryear{{Hailey-Dunsheath}, {Nikola}, {Stacey},
  {Oberst}, {Parshley}, {Bradford}, {Ade}  \& {Tucker}}{{Hailey-Dunsheath}
  et~al.}{2008}]{hd08}
{Hailey-Dunsheath} S.,  {Nikola} T.,  {Stacey} G.~J.,  {Oberst} T.~E.,
  {Parshley} S.~C.,  {Bradford} C.~M.,  {Ade} P.~A.~R.,   {Tucker} C.~E.,
  2008, \mn@doi [\apj] {10.1086/595840}, \href
  {https://ui.adsabs.harvard.edu/abs/2008ApJ...689L.109H} {689, L109}

\bibitem[\protect\citeauthoryear{{Hailey-Dunsheath}, {Nikola}, {Stacey},
  {Oberst}, {Parshley}, {Benford}, {Staguhn}  \& {Tucker}}{{Hailey-Dunsheath}
  et~al.}{2010}]{hd10}
{Hailey-Dunsheath} S.,  {Nikola} T.,  {Stacey} G.~J.,  {Oberst} T.~E.,
  {Parshley} S.~C.,  {Benford} D.~J.,  {Staguhn} J.~G.,   {Tucker} C.~E.,
  2010, \mn@doi [\apj] {10.1088/2041-8205/714/1/L162}, \href
  {https://ui.adsabs.harvard.edu/abs/2010ApJ...714L.162H} {714, L162}

\bibitem[\protect\citeauthoryear{{Harrington} et~al.,}{{Harrington}
  et~al.}{2016}]{harrington16}
{Harrington} K.~C.,  et~al., 2016, \mn@doi [\mnras] {10.1093/mnras/stw614},
  \href {http://adsabs.harvard.edu/abs/2016MNRAS.458.4383H} {458, 4383}

\bibitem[\protect\citeauthoryear{{Harrington} et~al.,}{{Harrington}
  et~al.}{2018}]{harrington18}
{Harrington} K.~C.,  et~al., 2018, \mn@doi [\mnras] {10.1093/mnras/stx3043},
  \href {https://ui.adsabs.harvard.edu/#abs/2018MNRAS.474.3866H} {474, 3866}

\bibitem[\protect\citeauthoryear{{Helou}, {Soifer}  \&
  {Rowan-Robinson}}{{Helou} et~al.}{1985}]{Helou1985}
{Helou} G.,  {Soifer} B.~T.,   {Rowan-Robinson} M.,  1985, \mn@doi [\apjl]
  {10.1086/184556}, \href {http://adsabs.harvard.edu/abs/1985ApJ...298L...7H}
  {298, L7}

\bibitem[\protect\citeauthoryear{{Herrera-Camus} et~al.,}{{Herrera-Camus}
  et~al.}{2016}]{hc16}
{Herrera-Camus} R.,  et~al., 2016, \mn@doi [\apj]
  {10.3847/0004-637X/826/2/175}, \href
  {https://ui.adsabs.harvard.edu/#abs/2016ApJ...826..175H} {826, 175}

\bibitem[\protect\citeauthoryear{{Heyminck}, {Kasemann}, {G{\"u}sten}, {de
  Lange}  \& {Graf}}{{Heyminck} et~al.}{2006}]{heyminck06}
{Heyminck} S.,  {Kasemann} C.,  {G{\"u}sten} R.,  {de Lange} G.,   {Graf}
  U.~U.,  2006, \mn@doi [\aap] {10.1051/0004-6361:20065413}, \href
  {https://ui.adsabs.harvard.edu/#abs/2006A&A...454L..21H} {454, L21}

\bibitem[\protect\citeauthoryear{{Hickox} \& {Alexander}}{{Hickox} \&
  {Alexander}}{2018}]{hickox18}
{Hickox} R.~C.,  {Alexander} D.~M.,  2018, \mn@doi [Annual Review of Astronomy
  and Astrophysics] {10.1146/annurev-astro-081817-051803}, \href
  {https://ui.adsabs.harvard.edu/\#abs/2018ARA&A..56..625H} {56, 625}

\bibitem[\protect\citeauthoryear{{Hopkins}, {Hernquist}, {Cox}  \&
  {Kere{\v{s}}}}{{Hopkins} et~al.}{2008}]{hopkins08}
{Hopkins} P.~F.,  {Hernquist} L.,  {Cox} T.~J.,   {Kere{\v{s}}} D.,  2008,
  \mn@doi [The Astrophysical Journal Supplement Series] {10.1086/524362}, \href
  {https://ui.adsabs.harvard.edu/#abs/2008ApJS..175..356H} {175, 356}

\bibitem[\protect\citeauthoryear{{Kamenetzky}, {Rangwala}, {Glenn}, {Maloney}
  \& {Conley}}{{Kamenetzky} et~al.}{2016}]{kamenetzky16}
{Kamenetzky} J.,  {Rangwala} N.,  {Glenn} J.,  {Maloney} P.~R.,   {Conley} A.,
  2016, \mn@doi [\apj] {10.3847/0004-637X/829/2/93}, \href
  {https://ui.adsabs.harvard.edu/#abs/2016ApJ...829...93K} {829, 93}

\bibitem[\protect\citeauthoryear{{Kennicutt}}{{Kennicutt}}{1998}]{kennicutt98}
{Kennicutt} J. R.~C.,  1998, \mn@doi [\araa] {10.1146/annurev.astro.36.1.189},
  \href {http://adsabs.harvard.edu/abs/1998ARA\&A..36..189K} {36, 189}

\bibitem[\protect\citeauthoryear{{Kennicutt} \& {Evans}}{{Kennicutt} \&
  {Evans}}{2012}]{kennicutt12}
{Kennicutt} R.~C.,  {Evans} N.~J.,  2012, \mn@doi [Annual Review of Astronomy
  and Astrophysics] {10.1146/annurev-astro-081811-125610}, \href
  {https://ui.adsabs.harvard.edu/#abs/2012ARA&A..50..531K} {50, 531}

\bibitem[\protect\citeauthoryear{{Kirkpatrick} et~al.,}{{Kirkpatrick}
  et~al.}{2017}]{kirkpatrick17}
{Kirkpatrick} A.,  et~al., 2017, \mn@doi [\apj] {10.3847/1538-4357/aa911d},
  \href {https://ui.adsabs.harvard.edu/#abs/2017ApJ...849..111K} {849, 111}

\bibitem[\protect\citeauthoryear{{Klein}, {Philipp}, {Kr{\"a}mer}, {Kasemann},
  {G{\"u}sten}  \& {Menten}}{{Klein} et~al.}{2006}]{klein06}
{Klein} B.,  {Philipp} S.~D.,  {Kr{\"a}mer} I.,  {Kasemann} C.,  {G{\"u}sten}
  R.,   {Menten} K.~M.,  2006, \mn@doi [\aap] {10.1051/0004-6361:20065415},
  \href {https://ui.adsabs.harvard.edu/#abs/2006A&A...454L..29K} {454, L29}

\bibitem[\protect\citeauthoryear{{Kruijssen} \& {Longmore}}{{Kruijssen} \&
  {Longmore}}{2013}]{kruijssen13}
{Kruijssen} J.~M.~D.,  {Longmore} S.~N.,  2013, \mn@doi [\mnras]
  {10.1093/mnras/stt1634}, \href
  {http://adsabs.harvard.edu/abs/2013MNRAS.435.2598K} {435, 2598}

\bibitem[\protect\citeauthoryear{{Lamarche} et~al.,}{{Lamarche}
  et~al.}{2017}]{lamarche17}
{Lamarche} C.,  et~al., 2017, \mn@doi [\apj] {10.3847/1538-4357/836/1/123},
  \href {https://ui.adsabs.harvard.edu/#abs/2017ApJ...836..123L} {836, 123}

\bibitem[\protect\citeauthoryear{{Lamarche} et~al.,}{{Lamarche}
  et~al.}{2018}]{lamarche18}
{Lamarche} C.,  et~al., 2018, preprint, \href
  {https://ui.adsabs.harvard.edu/#abs/2018arXiv180909630L} {p.
  arXiv:1809.09630} (\mn@eprint {arXiv} {1809.09630})

\bibitem[\protect\citeauthoryear{{Leung} et~al.,}{{Leung}
  et~al.}{2019}]{leung19}
{Leung} T.~K.~D.,  et~al., 2019, \mn@doi [\apj] {10.3847/1538-4357/aaf860},
  \href {https://ui.adsabs.harvard.edu/\#abs/2019ApJ...871...85L} {871, 85}

\bibitem[\protect\citeauthoryear{{Lu} et~al.,}{{Lu} et~al.}{2017}]{lu17}
{Lu} N.,  et~al., 2017, \mn@doi [\apjs] {10.3847/1538-4365/aa6476}, \href
  {http://adsabs.harvard.edu/abs/2017ApJS..230....1L} {230, 1}

\bibitem[\protect\citeauthoryear{{Lu} et~al.,}{{Lu} et~al.}{2018}]{lu18}
{Lu} N.,  et~al., 2018, \mn@doi [\apj] {10.3847/1538-4357/aad3c9}, \href
  {https://ui.adsabs.harvard.edu/#abs/2018ApJ...864...38L} {864, 38}

\bibitem[\protect\citeauthoryear{{Madau and Dickinson,}}{{Madau and
  Dickinson,}}{2014}]{madau14}
{Madau and Dickinson,} 2014, \mn@doi [\araa]
  {10.1146/annurev-astro-081811-125615}, \href
  {http://adsabs.harvard.edu/abs/2014ARA\%26A..52..415M} {52, 415}

\bibitem[\protect\citeauthoryear{{Maeder} \& {Meynet}}{{Maeder} \&
  {Meynet}}{2000}]{maeder00}
{Maeder} A.,  {Meynet} G.,  2000, \aap, \href
  {https://ui.adsabs.harvard.edu/#abs/2000A&A...361..159M} {361, 159}

\bibitem[\protect\citeauthoryear{{Magdis} et~al.,}{{Magdis}
  et~al.}{2016}]{magdis16}
{Magdis} G.~E.,  et~al., 2016, \mn@doi [\mnras] {10.1093/mnras/stv2931}, \href
  {https://ui.adsabs.harvard.edu/#abs/2016MNRAS.456.4533M} {456, 4533}

\bibitem[\protect\citeauthoryear{{Magnelli} et~al.,}{{Magnelli}
  et~al.}{2014}]{Magnelli2014}
{Magnelli} B.,  et~al., 2014, \mn@doi [\aap] {10.1051/0004-6361/201322217},
  \href {http://adsabs.harvard.edu/abs/2014A%26A...561A..86M} {561, A86}

\bibitem[\protect\citeauthoryear{{Maiolino} et~al.,}{{Maiolino}
  et~al.}{2005}]{maoilino05}
{Maiolino} R.,  et~al., 2005, \mn@doi [\aap] {10.1051/0004-6361:200500165},
  \href {https://ui.adsabs.harvard.edu/#abs/2005A&A...440L..51M} {440, L51}

\bibitem[\protect\citeauthoryear{{Maiolino}, {Caselli}, {Nagao}, {Walmsley},
  {De Breuck}  \& {Meneghetti}}{{Maiolino} et~al.}{2009}]{Maiolino2009}
{Maiolino} R.,  {Caselli} P.,  {Nagao} T.,  {Walmsley} M.,  {De Breuck} C.,
  {Meneghetti} M.,  2009, \mn@doi [\aap] {10.1051/0004-6361/200912265}, \href
  {http://adsabs.harvard.edu/abs/2009A%26A...500L...1M} {500, L1}

\bibitem[\protect\citeauthoryear{{Malhotra} et~al.,}{{Malhotra}
  et~al.}{2001}]{malhotra01}
{Malhotra} S.,  et~al., 2001, \mn@doi [\apj] {10.1086/323046}, \href
  {https://ui.adsabs.harvard.edu/abs/2001ApJ...561..766M} {561, 766}

\bibitem[\protect\citeauthoryear{{Marganian}, {Garwood}, {Braatz}, {Radziwill}
  \& {Maddalena}}{{Marganian} et~al.}{2013}]{marganian13}
{Marganian} P.,  {Garwood} R.~W.,  {Braatz} J.~A.,  {Radziwill} N.~M.,
  {Maddalena} R.~J.,  2013, {GBTIDL: Reduction and Analysis of GBT Spectral
  Line Data}, Astrophysics Source Code Library (\mn@eprint {ascl} {1303.019})

\bibitem[\protect\citeauthoryear{{Marrone} et~al.,}{{Marrone}
  et~al.}{2018}]{marrone18}
{Marrone} D.~P.,  et~al., 2018, \mn@doi [\nat] {10.1038/nature24629}, \href
  {https://ui.adsabs.harvard.edu/#abs/2018Natur.553...51M} {553, 51}

\bibitem[\protect\citeauthoryear{{Marshall} et~al.,}{{Marshall}
  et~al.}{2016}]{marshall16}
{Marshall} P.~J.,  et~al., 2016, \mn@doi [\mnras] {10.1093/mnras/stv2009},
  \href {https://ui.adsabs.harvard.edu/#abs/2016MNRAS.455.1171M} {455, 1171}

\bibitem[\protect\citeauthoryear{{Mashian} et~al.,}{{Mashian}
  et~al.}{2015}]{mashian15}
{Mashian} N.,  et~al., 2015, \mn@doi [\apj] {10.1088/0004-637X/802/2/81}, \href
  {https://ui.adsabs.harvard.edu/#abs/2015ApJ...802...81M} {802, 81}

\bibitem[\protect\citeauthoryear{{McKee} \& {Williams}}{{McKee} \&
  {Williams}}{1997}]{mckee97}
{McKee} C.~F.,  {Williams} J.~P.,  1997, \mn@doi [\apj] {10.1086/303587}, \href
  {https://ui.adsabs.harvard.edu/#abs/1997ApJ...476..144M} {476, 144}

\bibitem[\protect\citeauthoryear{{McPartland}, {Sanders}, {Kewley}  \&
  {Leslie}}{{McPartland} et~al.}{2019}]{mcpartland19}
{McPartland} C.,  {Sanders} D.~B.,  {Kewley} L.~J.,   {Leslie} S.~K.,  2019,
  \mn@doi [\mnras] {10.1093/mnrasl/sly202}, \href
  {https://ui.adsabs.harvard.edu/#abs/2019MNRAS.482L.129M} {482, L129}

\bibitem[\protect\citeauthoryear{{Moser} et~al.,}{{Moser}
  et~al.}{2017}]{moser17}
{Moser} L.,  et~al., 2017, \mn@doi [\aap] {10.1051/0004-6361/201628385}, \href
  {http://adsabs.harvard.edu/abs/2017A%26A...603A..68M} {603, A68}

\bibitem[\protect\citeauthoryear{{Nagao}, {Maiolino}, {De Breuck}, {Caselli},
  {Hatsukade}  \& {Saigo}}{{Nagao} et~al.}{2012}]{nagao12}
{Nagao} T.,  {Maiolino} R.,  {De Breuck} C.,  {Caselli} P.,  {Hatsukade} B.,
  {Saigo} K.,  2012, \mn@doi [\aap] {10.1051/0004-6361/201219518}, \href
  {https://ui.adsabs.harvard.edu/#abs/2012A&A...542L..34N} {542, L34}

\bibitem[\protect\citeauthoryear{{Nayyeri} et~al.,}{{Nayyeri}
  et~al.}{2016}]{nayyeri16}
{Nayyeri} H.,  et~al., 2016, \mn@doi [\apj] {10.3847/0004-637X/823/1/17}, \href
  {https://ui.adsabs.harvard.edu/#abs/2016ApJ...823...17N} {823, 17}

\bibitem[\protect\citeauthoryear{{Negrello}, {Hopwood}, {De Zotti}, {Cooray},
  {Verma}  et~al.}{{Negrello} et~al.}{2010}]{negrello10}
{Negrello} M.,  {Hopwood} R.,  {De Zotti} G.,  {Cooray} A.,  {Verma} A.,
  et~al., 2010, \mn@doi [Science] {10.1126/science.1193420}, \href
  {http://adsabs.harvard.edu/abs/2010Sci...330..800N} {330, 800}

\bibitem[\protect\citeauthoryear{{Nikola}, {Stacey}, {Brisbin}, {Ferkinhoff},
  {Hailey-Dunsheath}, {Parshley}  \& {Tucker}}{{Nikola}
  et~al.}{2011}]{nikola11}
{Nikola} T.,  {Stacey} G.~J.,  {Brisbin} D.,  {Ferkinhoff} C.,
  {Hailey-Dunsheath} S.,  {Parshley} S.,   {Tucker} C.,  2011, \mn@doi [\apj]
  {10.1088/0004-637X/742/2/88}, \href
  {https://ui.adsabs.harvard.edu/abs/2011ApJ...742...88N} {742, 88}

\bibitem[\protect\citeauthoryear{{Oberst} et~al.,}{{Oberst}
  et~al.}{2006}]{oberst06}
{Oberst} T.~E.,  et~al., 2006, \mn@doi [\apj] {10.1086/510289}, \href
  {https://ui.adsabs.harvard.edu/#abs/2006ApJ...652L.125O} {652, L125}

\bibitem[\protect\citeauthoryear{{Oberst}, {Parshley}, {Nikola}, {Stacey},
  {L{\"o}hr}, {Lane}, {Stark}  \& {Kamenetzky}}{{Oberst}
  et~al.}{2011}]{oberst11}
{Oberst} T.~E.,  {Parshley} S.~C.,  {Nikola} T.,  {Stacey} G.~J.,  {L{\"o}hr}
  A.,  {Lane} A.~P.,  {Stark} A.~A.,   {Kamenetzky} J.,  2011, \mn@doi [\apj]
  {10.1088/0004-637X/739/2/100}, \href
  {https://ui.adsabs.harvard.edu/#abs/2011ApJ...739..100O} {739, 100}

\bibitem[\protect\citeauthoryear{{Panuzzo} et~al.,}{{Panuzzo}
  et~al.}{2010}]{panuzzo10}
{Panuzzo} P.,  et~al., 2010, \mn@doi [\aap] {10.1051/0004-6361/201014558},
  \href {https://ui.adsabs.harvard.edu/abs/2010A&A...518L..37P} {518, L37}

\bibitem[\protect\citeauthoryear{{Papadopoulos}, {van der Werf}, {Xilouris},
  {Isaak}, {Gao}  \& {M{\"u}hle}}{{Papadopoulos} et~al.}{2012}]{pap12}
{Papadopoulos} P.~P.,  {van der Werf} P.~P.,  {Xilouris} E.~M.,  {Isaak} K.~G.,
   {Gao} Y.,   {M{\"u}hle} S.,  2012, \mn@doi [\mnras]
  {10.1111/j.1365-2966.2012.21001.x}, \href
  {http://adsabs.harvard.edu/abs/2012MNRAS.426.2601P} {426, 2601}

\bibitem[\protect\citeauthoryear{{Parkin} et~al.,}{{Parkin}
  et~al.}{2013}]{parkin13}
{Parkin} T.~J.,  et~al., 2013, \mn@doi [\apj] {10.1088/0004-637X/776/2/65},
  \href {https://ui.adsabs.harvard.edu/#abs/2013ApJ...776...65P} {776, 65}

\bibitem[\protect\citeauthoryear{{Pavesi} et~al.,}{{Pavesi}
  et~al.}{2016}]{pavesi16}
{Pavesi} R.,  et~al., 2016, \mn@doi [\apj] {10.3847/0004-637X/832/2/151}, \href
  {http://adsabs.harvard.edu/abs/2016ApJ...832..151P} {832, 151}

\bibitem[\protect\citeauthoryear{{Pavesi}, {Riechers}, {Faisst}, {Stacey}  \&
  {Capak}}{{Pavesi} et~al.}{2018a}]{pavesi18c}
{Pavesi} R.,  {Riechers} D.~A.,  {Faisst} A.~L.,  {Stacey} G.~J.,   {Capak}
  P.~L.,  2018a, preprint, \href
  {https://ui.adsabs.harvard.edu/#abs/2018arXiv181200006P} {p.
  arXiv:1812.00006} (\mn@eprint {arXiv} {1812.00006})

\bibitem[\protect\citeauthoryear{{Pavesi} et~al.,}{{Pavesi}
  et~al.}{2018b}]{pavesi18b}
{Pavesi} R.,  et~al., 2018b, \mn@doi [\apj] {10.3847/1538-4357/aac6b6}, \href
  {https://ui.adsabs.harvard.edu/#abs/2018ApJ...861...43P} {861, 43}

\bibitem[\protect\citeauthoryear{{Petuchowski}, {Bennett}, {Haas}, {Erickson},
  {Lord}, {Rubin}, {Colgan}  \& {Hollenbach}}{{Petuchowski}
  et~al.}{1994}]{petuchowski94}
{Petuchowski} S.~J.,  {Bennett} C.~L.,  {Haas} M.~R.,  {Erickson} E.~F.,
  {Lord} S.~D.,  {Rubin} R.~H.,  {Colgan} S. W.~J.,   {Hollenbach} D.~J.,
  1994, \mn@doi [\apj] {10.1086/187354}, \href
  {https://ui.adsabs.harvard.edu/#abs/1994ApJ...427L..17P} {427, L17}

\bibitem[\protect\citeauthoryear{{Planck Collaboration XXVII}}{{Planck
  Collaboration XXVII}}{2015}]{planck26}
{Planck Collaboration XXVII} 2015, \mn@doi [\aap]
  {10.1051/0004-6361/201424790}, \href
  {http://adsabs.harvard.edu/abs/2015A26A...582A..30P} {582, A30}

\bibitem[\protect\citeauthoryear{{Puls}, {Vink}  \& {Najarro}}{{Puls}
  et~al.}{2008}]{puls08}
{Puls} J.,  {Vink} J.~S.,   {Najarro} F.,  2008, \mn@doi [Astronomy and
  Astrophysics Review] {10.1007/s00159-008-0015-8}, \href
  {https://ui.adsabs.harvard.edu/#abs/2008A&ARv..16..209P} {16, 209}

\bibitem[\protect\citeauthoryear{{Rawle} et~al.,}{{Rawle}
  et~al.}{2014}]{rawle14}
{Rawle} T.~D.,  et~al., 2014, \mn@doi [\apj] {10.1088/0004-637X/783/1/59},
  \href {https://ui.adsabs.harvard.edu/#abs/2014ApJ...783...59R} {783, 59}

\bibitem[\protect\citeauthoryear{{Riechers} et~al.,}{{Riechers}
  et~al.}{2013a}]{riechers13}
{Riechers} D.~A.,  et~al., 2013a, \mn@doi [\nat] {10.1038/nature12050}, \href
  {http://adsabs.harvard.edu/abs/2013Natur.496..329R} {496, 329}

\bibitem[\protect\citeauthoryear{{Riechers} et~al.,}{{Riechers}
  et~al.}{2013b}]{Riechers2013Nature}
{Riechers} D.~A.,  et~al., 2013b, \mn@doi [\nat] {10.1038/nature12050}, \href
  {http://adsabs.harvard.edu/abs/2013Natur.496..329R} {496, 329}

\bibitem[\protect\citeauthoryear{{Riechers} et~al.,}{{Riechers}
  et~al.}{2014}]{riechers14}
{Riechers} D.~A.,  et~al., 2014, \mn@doi [\apj] {10.1088/0004-637X/796/2/84},
  \href {https://ui.adsabs.harvard.edu/#abs/2014ApJ...796...84R} {796, 84}

\bibitem[\protect\citeauthoryear{{Rivera} et~al.,}{{Rivera}
  et~al.}{2018}]{rivera18}
{Rivera} J.,  et~al., 2018, preprint, \href
  {https://ui.adsabs.harvard.edu/#abs/2018arXiv180708895R} {p.
  arXiv:1807.08895} (\mn@eprint {arXiv} {1807.08895})

\bibitem[\protect\citeauthoryear{{Rosdahl}, {Schaye}, {Dubois}, {Kimm}  \&
  {Teyssier}}{{Rosdahl} et~al.}{2017}]{rosdahl17}
{Rosdahl} J.,  {Schaye} J.,  {Dubois} Y.,  {Kimm} T.,   {Teyssier} R.,  2017,
  \mn@doi [\mnras] {10.1093/mnras/stw3034}, \href
  {https://ui.adsabs.harvard.edu/\#abs/2017MNRAS.466...11R} {466, 11}

\bibitem[\protect\citeauthoryear{{Rosenberg} et~al.,}{{Rosenberg}
  et~al.}{2015}]{rosenberg15}
{Rosenberg} M.~J.~F.,  et~al., 2015, \mn@doi [\apj]
  {10.1088/0004-637X/801/2/72}, \href
  {http://adsabs.harvard.edu/abs/2015ApJ...801...72R} {801, 72}

\bibitem[\protect\citeauthoryear{{R{\'o}{\.z}a{\'n}ska}, {Czerny},
  {Kunneriath}, {Adhikari}, {Karas}  \&
  {Mo{\'s}cibrodzka}}{{R{\'o}{\.z}a{\'n}ska} et~al.}{2014}]{rozanska14}
{R{\'o}{\.z}a{\'n}ska} A.,  {Czerny} B.,  {Kunneriath} D.,  {Adhikari} T.~P.,
  {Karas} V.,   {Mo{\'s}cibrodzka} M.,  2014, \mn@doi [\mnras]
  {10.1093/mnras/stu2066}, \href
  {http://adsabs.harvard.edu/abs/2014MNRAS.445.4385R} {445, 4385}

\bibitem[\protect\citeauthoryear{{R{\'o}{\.z}a{\'n}ska}, {Kunneriath},
  {Czerny}, {Adhikari}  \& {Karas}}{{R{\'o}{\.z}a{\'n}ska}
  et~al.}{2017}]{rozanska17}
{R{\'o}{\.z}a{\'n}ska} A.,  {Kunneriath} D.,  {Czerny} B.,  {Adhikari} T.~P.,
  {Karas} V.,  2017, \mn@doi [\mnras] {10.1093/mnras/stw2460}, \href
  {http://adsabs.harvard.edu/abs/2017MNRAS.464.2090R} {464, 2090}

\bibitem[\protect\citeauthoryear{{Rujopakarn} et~al.,}{{Rujopakarn}
  et~al.}{2016}]{rujo16}
{Rujopakarn} W.,  et~al., 2016, \mn@doi [\apj] {10.3847/0004-637X/833/1/12},
  \href {https://ui.adsabs.harvard.edu/#abs/2016ApJ...833...12R} {833, 12}

\bibitem[\protect\citeauthoryear{{Salom{\'e}}, {Gu{\'e}lin}, {Downes}, {Cox},
  {Guilloteau}, {Omont}, {Gavazzi}  \& {Neri}}{{Salom{\'e}}
  et~al.}{2012}]{salome12}
{Salom{\'e}} P.,  {Gu{\'e}lin} M.,  {Downes} D.,  {Cox} P.,  {Guilloteau} S.,
  {Omont} A.,  {Gavazzi} R.,   {Neri} R.,  2012, \mn@doi [\aap]
  {10.1051/0004-6361/201219955}, \href
  {https://ui.adsabs.harvard.edu/#abs/2012A&A...545A..57S} {545, A57}

\bibitem[\protect\citeauthoryear{{Sanders} \& {Mirabel}}{{Sanders} \&
  {Mirabel}}{1996}]{sanders96}
{Sanders} D.~B.,  {Mirabel} I.~F.,  1996, \mn@doi [\araa]
  {10.1146/annurev.astro.34.1.749}, \href
  {http://adsabs.harvard.edu/abs/1996ARA26A..34..749S} {34, 749}

\bibitem[\protect\citeauthoryear{{Savage} \& {Sembach}}{{Savage} \&
  {Sembach}}{1996}]{savage96}
{Savage} B.~D.,  {Sembach} K.~R.,  1996, \mn@doi [\apj] {10.1086/177919}, \href
  {https://ui.adsabs.harvard.edu/#abs/1996ApJ...470..893S} {470, 893}

\bibitem[\protect\citeauthoryear{{Scannapieco} et~al.,}{{Scannapieco}
  et~al.}{2012}]{scannapieco12}
{Scannapieco} C.,  et~al., 2012, \mn@doi [\mnras]
  {10.1111/j.1365-2966.2012.20993.x}, \href
  {https://ui.adsabs.harvard.edu/\#abs/2012MNRAS.423.1726S} {423, 1726}

\bibitem[\protect\citeauthoryear{{Schinnerer} et~al.,}{{Schinnerer}
  et~al.}{2016}]{Schinnerer2016}
{Schinnerer} E.,  et~al., 2016, \mn@doi [\apj] {10.3847/1538-4357/833/1/112},
  \href {http://adsabs.harvard.edu/abs/2016ApJ...833..112S} {833, 112}

\bibitem[\protect\citeauthoryear{{Schulz} et~al.,}{{Schulz}
  et~al.}{2017}]{Schulz17}
{Schulz} B.,  et~al., 2017, arXiv e-prints, \href
  {https://ui.adsabs.harvard.edu/\#abs/2017arXiv170600448S} {p.
  arXiv:1706.00448}

\bibitem[\protect\citeauthoryear{{Scoville} et~al.}{{Scoville}
  et~al.}{2014}]{scoville14}
{Scoville} N.,  et~al., 2014, \mn@doi [\apj] {10.1088/0004-637X/783/2/84},
  \href {http://adsabs.harvard.edu/abs/2014ApJ...783...84S} {783, 84}

\bibitem[\protect\citeauthoryear{{Scoville} et~al.,}{{Scoville}
  et~al.}{2016}]{scoville16}
{Scoville} N.,  et~al., 2016, \mn@doi [\apj] {10.3847/0004-637X/824/1/63},
  \href {http://adsabs.harvard.edu/abs/2016ApJ...824...63S} {824, 63}

\bibitem[\protect\citeauthoryear{{Scoville} et~al.,}{{Scoville}
  et~al.}{2017}]{scoville17}
{Scoville} N.,  et~al., 2017, \mn@doi [\apj] {10.3847/1538-4357/aa61a0}, \href
  {http://adsabs.harvard.edu/abs/2017ApJ...837..150S} {837, 150}

\bibitem[\protect\citeauthoryear{{Serjeant}}{{Serjeant}}{2012}]{serj12}
{Serjeant} S.,  2012, \mn@doi [\mnras] {10.1111/j.1365-2966.2012.20761.x},
  \href {http://adsabs.harvard.edu/abs/2012MNRAS.424.2429S} {424, 2429}

\bibitem[\protect\citeauthoryear{{Siebenmorgen}, {Kr{\"u}gel}  \&
  {Spoon}}{{Siebenmorgen} et~al.}{2004}]{siebenmorgen04}
{Siebenmorgen} R.,  {Kr{\"u}gel} E.,   {Spoon} H.~W.~W.,  2004, \mn@doi [\aap]
  {10.1051/0004-6361:20031633}, \href
  {https://ui.adsabs.harvard.edu/#abs/2004A&A...414..123S} {414, 123}

\bibitem[\protect\citeauthoryear{{Siebenmorgen}, {Heymann}  \&
  {Efstathiou}}{{Siebenmorgen} et~al.}{2015}]{siebenmorgen15}
{Siebenmorgen} R.,  {Heymann} F.,   {Efstathiou} A.,  2015, \mn@doi [\aap]
  {10.1051/0004-6361/201526034}, \href
  {https://ui.adsabs.harvard.edu/#abs/2015A&A...583A.120S} {583, A120}

\bibitem[\protect\citeauthoryear{{Solomon} \& {Vanden Bout}}{{Solomon} \&
  {Vanden Bout}}{2005a}]{sv05}
{Solomon} P.~M.,  {Vanden Bout} P.~A.,  2005a, \mn@doi [\araa]
  {10.1146/annurev.astro.43.051804.102221}, \href
  {http://adsabs.harvard.edu/abs/2005ARA\%26A..43..677S} {43, 677}

\bibitem[\protect\citeauthoryear{{Solomon} \& {Vanden Bout}}{{Solomon} \&
  {Vanden Bout}}{2005b}]{Solomon2005}
{Solomon} P.~M.,  {Vanden Bout} P.~A.,  2005b, \mn@doi [\araa]
  {10.1146/annurev.astro.43.051804.102221}, \href
  {http://adsabs.harvard.edu/abs/2005ARA%26A..43..677S} {43, 677}

\bibitem[\protect\citeauthoryear{{Solomon}, {Downes}, {Radford}  \&
  {Barrett}}{{Solomon} et~al.}{1997}]{solomon97}
{Solomon} P.~M.,  {Downes} D.,  {Radford} S.~J.~E.,   {Barrett} J.~W.,  1997,
  \mn@doi [\apj] {10.1086/303765}, \href
  {http://adsabs.harvard.edu/abs/1997ApJ...478..144S} {478, 144}

\bibitem[\protect\citeauthoryear{{Spinoglio}, {Pereira-Santaella}, {Dasyra},
  {Calzoletti}, {Malkan}, {Tommasin}  \& {Busquet}}{{Spinoglio}
  et~al.}{2015}]{spinoglio15}
{Spinoglio} L.,  {Pereira-Santaella} M.,  {Dasyra} K.~M.,  {Calzoletti} L.,
  {Malkan} M.~A.,  {Tommasin} S.,   {Busquet} G.,  2015, \mn@doi [\apj]
  {10.1088/0004-637X/799/1/21}, \href
  {https://ui.adsabs.harvard.edu/#abs/2015ApJ...799...21S} {799, 21}

\bibitem[\protect\citeauthoryear{{Stacey} et~al.,}{{Stacey}
  et~al.}{2018}]{stacey17}
{Stacey} H.~R.,  et~al., 2018, \mn@doi [\mnras] {10.1093/mnras/sty458}, \href
  {https://ui.adsabs.harvard.edu/#abs/2018MNRAS.476.5075S} {476, 5075}

\bibitem[\protect\citeauthoryear{{Stanway} \& {Eldridge}}{{Stanway} \&
  {Eldridge}}{2018}]{stanway18}
{Stanway} E.~R.,  {Eldridge} J.~J.,  2018, preprint, \href
  {http://adsabs.harvard.edu/abs/2018arXiv181103856S} {} (\mn@eprint {arXiv}
  {1811.03856})

\bibitem[\protect\citeauthoryear{{Su} et~al.,}{{Su} et~al.}{2017}]{su17}
{Su} T.,  et~al., 2017, \mn@doi [\mnras] {10.1093/mnras/stw2334}, \href
  {https://ui.adsabs.harvard.edu/#abs/2017MNRAS.464..968S} {464, 968}

\bibitem[\protect\citeauthoryear{{Swinbank} et~al.}{{Swinbank}
  et~al.}{2011}]{swinbank11}
{Swinbank} A.~M.,  et~al., 2011, \mn@doi [\apj] {10.1088/0004-637X/742/1/11},
  \href {http://adsabs.harvard.edu/abs/2011ApJ...742...11S} {742, 11}

\bibitem[\protect\citeauthoryear{{Tacconi} et~al.,}{{Tacconi}
  et~al.}{2010}]{tacconi10}
{Tacconi} L.~J.,  et~al., 2010, \mn@doi [\nat] {10.1038/nature08773}, \href
  {http://adsabs.harvard.edu/abs/2010Natur.463..781T} {463, 781}

\bibitem[\protect\citeauthoryear{{Tacconi} et~al.,}{{Tacconi}
  et~al.}{2018}]{Tacconi2018}
{Tacconi} L.~J.,  et~al., 2018, \mn@doi [\apj] {10.3847/1538-4357/aaa4b4},
  \href {http://adsabs.harvard.edu/abs/2018ApJ...853..179T} {853, 179}

\bibitem[\protect\citeauthoryear{{Tayal}}{{Tayal}}{2011}]{tayal11}
{Tayal} S.~S.,  2011, \mn@doi [The Astrophysical Journal Supplement Series]
  {10.1088/0067-0049/195/2/12}, \href
  {https://ui.adsabs.harvard.edu/#abs/2011ApJS..195...12T} {195, 12}

\bibitem[\protect\citeauthoryear{{Tenorio-Tagle}, {Silich},
  {Mart{\'{\i}}nez-Gonz{\'a}lez}, {Mu{\~n}oz-Tu{\~n}{\'o}n}, {Palou{\v s}}  \&
  {W{\"u}nsch}}{{Tenorio-Tagle} et~al.}{2013}]{tenoriotagle13}
{Tenorio-Tagle} G.,  {Silich} S.,  {Mart{\'{\i}}nez-Gonz{\'a}lez} S.,
  {Mu{\~n}oz-Tu{\~n}{\'o}n} C.,  {Palou{\v s}} J.,   {W{\"u}nsch} R.,  2013,
  \mn@doi [\apj] {10.1088/0004-637X/778/2/159}, \href
  {http://adsabs.harvard.edu/abs/2013ApJ...778..159T} {778, 159}

\bibitem[\protect\citeauthoryear{{Tsai} et~al.}{{Tsai} et~al.}{2015}]{tsai15}
{Tsai} C.-W.,  et~al., 2015, \mn@doi [\apj] {10.1088/0004-637X/805/2/90}, \href
  {http://adsabs.harvard.edu/abs/2015ApJ...805...90T} {805, 90}

\bibitem[\protect\citeauthoryear{{Uzgil}, {Bradford}, {Hailey-Dunsheath},
  {Maloney}  \& {Aguirre}}{{Uzgil} et~al.}{2016}]{uzgil16}
{Uzgil} B.~D.,  {Bradford} C.~M.,  {Hailey-Dunsheath} S.,  {Maloney} P.~R.,
  {Aguirre} J.~E.,  2016, \mn@doi [\apj] {10.3847/0004-637X/832/2/209}, \href
  {https://ui.adsabs.harvard.edu/#abs/2016ApJ...832..209U} {832, 209}

\bibitem[\protect\citeauthoryear{{Vishwas} et~al.,}{{Vishwas}
  et~al.}{2018}]{vishwas18}
{Vishwas} A.,  et~al., 2018, \mn@doi [\apj] {10.3847/1538-4357/aab354}, \href
  {https://ui.adsabs.harvard.edu/#abs/2018ApJ...856..174V} {856, 174}

\bibitem[\protect\citeauthoryear{{Walter}, {Wei{\ss}}, {Riechers}, {Carilli},
  {Bertoldi}, {Cox}  \& {Menten}}{{Walter} et~al.}{2009}]{walter09}
{Walter} F.,  {Wei{\ss}} A.,  {Riechers} D.~A.,  {Carilli} C.~L.,  {Bertoldi}
  F.,  {Cox} P.,   {Menten} K.~M.,  2009, \mn@doi [\apj]
  {10.1088/0004-637X/691/1/L1}, \href
  {https://ui.adsabs.harvard.edu/#abs/2009ApJ...691L...1W} {691, L1}

\bibitem[\protect\citeauthoryear{{Wardlow} et~al.}{{Wardlow}
  et~al.}{2013}]{wardlow13}
{Wardlow} J.~L.,  et~al., 2013, \mn@doi [\apj] {10.1088/0004-637X/762/1/59},
  \href {http://adsabs.harvard.edu/abs/2013ApJ...762...59W} {762, 59}

\bibitem[\protect\citeauthoryear{{Wei{\ss}}, {Downes}, {Neri}, {Walter},
  {Henkel}, {Wilner}, {Wagg}  \& {Wiklind}}{{Wei{\ss}} et~al.}{2007}]{weiss07}
{Wei{\ss}} A.,  {Downes} D.,  {Neri} R.,  {Walter} F.,  {Henkel} C.,  {Wilner}
  D.~J.,  {Wagg} J.,   {Wiklind} T.,  2007, \mn@doi [\aap]
  {10.1051/0004-6361:20066117}, \href
  {https://ui.adsabs.harvard.edu/#abs/2007A&A...467..955W} {467, 955}

\bibitem[\protect\citeauthoryear{{Wei{\ss}}, {Kov{\'a}cs}, {G{\"u}sten},
  {Menten}, {Schuller}, {Siringo}  \& {Kreysa}}{{Wei{\ss}}
  et~al.}{2008}]{weiss08}
{Wei{\ss}} A.,  {Kov{\'a}cs} A.,  {G{\"u}sten} R.,  {Menten} K.~M.,  {Schuller}
  F.,  {Siringo} G.,   {Kreysa} E.,  2008, \mn@doi [\aap]
  {10.1051/0004-6361:200809909}, \href
  {https://ui.adsabs.harvard.edu/#abs/2008A&A...490...77W} {490, 77}

\bibitem[\protect\citeauthoryear{{Whitaker}, {Pope}, {Cybulski}, {Casey},
  {Popping}  \& {Yun}}{{Whitaker} et~al.}{2017}]{whit17}
{Whitaker} K.~E.,  {Pope} A.,  {Cybulski} R.,  {Casey} C.~M.,  {Popping} G.,
  {Yun} M.~S.,  2017, \mn@doi [\apj] {10.3847/1538-4357/aa94ce}, \href
  {https://ui.adsabs.harvard.edu/#abs/2017ApJ...850..208W} {850, 208}

\bibitem[\protect\citeauthoryear{{Zanella} et~al.,}{{Zanella}
  et~al.}{2018}]{zanella18}
{Zanella} A.,  et~al., 2018, \mn@doi [\mnras] {10.1093/mnras/sty2394}, \href
  {https://ui.adsabs.harvard.edu/#abs/2018MNRAS.481.1976Z} {481, 1976}

\bibitem[\protect\citeauthoryear{{Zavala} et~al.,}{{Zavala}
  et~al.}{2018}]{zavala18}
{Zavala} J.~A.,  et~al., 2018, \mn@doi [Nature Astronomy]
  {10.1038/s41550-017-0297-8}, \href
  {https://ui.adsabs.harvard.edu/#abs/2018NatAs...2...56Z} {2, 56}

\bibitem[\protect\citeauthoryear{{Zhang} et~al.,}{{Zhang}
  et~al.}{2018}]{zhang18}
{Zhang} Z.-Y.,  et~al., 2018, \mn@doi [\mnras] {10.1093/mnras/sty2082}, \href
  {https://ui.adsabs.harvard.edu/#abs/2018MNRAS.481...59Z} {481, 59}

\bibitem[\protect\citeauthoryear{{Zhao} et~al.,}{{Zhao} et~al.}{2013}]{zhao13}
{Zhao} Y.,  et~al., 2013, \mn@doi [\apjl] {10.1088/2041-8205/765/1/L13}, \href
  {http://adsabs.harvard.edu/abs/2013ApJ...765L..13Z} {765, L13}

\bibitem[\protect\citeauthoryear{{Zhao} et~al.,}{{Zhao} et~al.}{2016}]{zhao16}
{Zhao} Y.,  et~al., 2016, \mn@doi [\apj] {10.3847/0004-637X/819/1/69}, \href
  {https://ui.adsabs.harvard.edu/#abs/2016ApJ...819...69Z} {819, 69}

\makeatother
\end{thebibliography}


\bsp	
\label{lastpage}
\end{document}